\documentclass[amssymb,usenatbib,useAMS,usedcolumns]{mn2e}
\usepackage{times} 
\usepackage{latexsym,amssymb,amsmath,amsfonts}
\usepackage{rotating} 
\usepackage{float}
\usepackage{graphicx}
\usepackage{natbib}
\usepackage{longtable}
\usepackage{url}
\usepackage{dcolumn}
\usepackage{textcomp}
\begin{document}
\newcommand{\hon}{\mbox{H\,{\sc i}}}
\newcommand{\cat}{\mbox{Ca\,{\sc ii}}}
\newcommand{\nao}{\mbox{Na\,{\sc i}}}
\newcommand{\lapp}{\mbox{\raisebox{-0.3em}{$\stackrel{\textstyle <}{\sim}$}}}
\newcommand{\gapp}{\mbox{\raisebox{-0.3em}{$\stackrel{\textstyle >}{\sim}$}}}
\def\h2{$\rm H_2$}
\def\Nh2{$N$(H${_2}$)}
\def\chin{$\chi^2_{\nu}$}
\def\chiu{$\chi_{\rm UV}$}
\def\sys{J0041$-$0143~}
\def\lya{\ensuremath{{\rm Ly}\alpha~}}
\def\lymana{\ensuremath{{\rm Lyman}-\alpha~}}
\def\kms{km\,s$^{-1}$}
\def\cms{cm$^{-2}$}
\def\cc{cm$^{-3}$}
\def\zabs{$z_{\rm abs~}$}
\def\zem{$z_{\rm em~}$}
\def\nhi{$N$($\hon$)~}
\def\ln{log~$N$}
\def\nh{$n_{\rm H}$}
\def\ne{$n_{e}$}
\def\21{21-cm}
\def\ts{T$_{s}$}
\def\th{T$_{01}$}
\def\ll{$\lambda\lambda$}
\def\l{$\lambda$}
\def\fc{$f_{c}$}
\def\mjb{mJy~beam$^{-1}$~}
\def\taudv{$\int\tau dv$~}
\def\gal{UGC~00439~}
%
%
\title[Structures in extended \hon\ disc of UGC~00439]{
{Mapping kiloparsec-scale structures in the extended \hon\ disc of the galaxy UGC~000439 by \hon\ \21 absorption}  
\author[Dutta et al.]{R. Dutta$^1$\thanks{E-mail: rdutta@iucaa.in}, N. Gupta$^1$, R. Srianand$^1$, J. M. O'Meara$^2$ \\ 
$^1$ Inter-University Centre for Astronomy and Astrophysics, Post Bag 4, Ganeshkhind, Pune 411007, India \\
$^2$ Department of Chemistry and Physics, Saint Michael's College, One Winooski Park, Colchester, VT 05439, USA
} 
}
\date{Accepted. Received; in original form }
\maketitle
\label{firstpage}
%
%
\begin {abstract}  
\par\noindent
We study the properties of H~{\sc i} gas in the outer regions ($\sim$2$r_{25}$) of a spiral galaxy, UGC\,00439 
($z$ = 0.01769), using \hon\ \21 absorption towards different components of an extended background radio source, 
J0041$-$0043 ($z$ = 1.679). The radio source exhibits a compact core coincident with the optical quasar and two lobes 
separated by $\sim$7~kpc, all at an impact parameter $\sim$25\,kpc. The \hon\ \21 absorption detected towards the 
southern lobe is found to extend over $\sim$2\,kpc$^2$. The absorbing gas shows sub-kpc-scale structures with the 
line-of-sight velocities dominated by turbulent motions. Much larger optical depth variations over 4-7\,kpc-scale 
are revealed by the non-detection of \hon\ \21 absorption towards the radio core and the northern lobe, and the 
detection of Na~{\sc i} and Ca~{\sc ii} absorption towards the quasar. This could reflect a patchy distribution of 
cold gas in the extended \hon\ disc. We also detect H~{\sc i} \21 emission from UGC~00439 and two other galaxies 
within $\sim$150\,kpc to it, that probably form an interacting group. However, no  H~{\sc i} \21 emission from 
the absorbing gas is detected. Assuming a linear extent of $\sim$4\,kpc, as required to cover both the core and 
the southern lobe, we constrain the spin temperature $\lesssim 300$\,K for the absorbing gas. The kinematics of 
the gas and the lack of signatures of any ongoing {\it in situ} star formation are consistent with the absorbing gas 
being at the kinematical minor axis and corotating with the galaxy. Deeper H~{\sc i} \21 observations would help 
to map in greater detail both the large- and small-scale structures in the H~{\sc i} gas associated with UGC\,00439. 
\end {abstract}  
%
%
%
%
\section{Introduction} 
\label{sec_introduction}  
In multiphase pressure equilibrium models, as well as supernova driven hydrodynamical simulations, the physical conditions, 
volume filling factors and pc- to kpc-scale structures of different phases of $\hon$ gas in the interstellar medium (ISM) of 
galaxies are found to depend on various radiative and mechanical feedback processes associated with the {\it in situ} star formation 
\citep{mckee1977,wolfire1995,deavillez2004,gent2013,gatto2015}. Therefore, understanding the properties of \hon\ gas phases, 
especially the cold neutral medium (CNM) which is closely related to the molecular gas, is crucial to understand processes 
through which gas gets converted to stars. 

In our Galaxy, \hon\ \21 emission and absorption line observations have been extensively used to characterize physical conditions 
and pc- to kpc-scale structures in the ISM \citep[][]{heiles84, Frail94, Deshpande00, heiles2003, Brogan05}. Since \hon\ \21 
absorption is sensitive to the spin temperature ($T_s$), which is coupled to the kinetic temperature of the gas \citep{roy2006}, 
repeated measurements of \hon\ \21 absorption towards high-velocity pulsars, extended radio sources, and radio sources with jets having 
proper motion, have been particularly useful in revealing small-scale structures of CNM. However, due to the scarcity of known strong 
and extended radio sources at close impact parameters with respect to galaxies, it has been possible to extend such studies to very
few external galaxies \citep[see][]{wolfe1982,briggs2001,kanekar2001,kanekar2005,srianand2013}. \hon\ \21 emission line observations of 
nearby dwarf and spiral galaxies show that the properties of CNM and warm neutral medium (WNM) and $\sim$100\,pc- to 2\,kpc-scale structures 
detected in the \hon\ gas are coupled to the local star formation in galaxies \citep[e.g.][]{Tamburro2009, Bagetakos11, Ianjamasimanana12}. 
However, in the absence of absorption line measurements it is not known whether the \hon\ \21 emission line components exhibiting 
smaller line widths (and hence assumed to correspond to CNM) are truly cold. Therefore, the contributions due to turbulent motions 
and the processes driving the observed properties of \hon\ gas are poorly constrained.  

In this paper, we present results from $\hon$ \21 absorption and emission line observations of a low-$z$ Quasar-galaxy-pair (QGP)\footnote{Quasar 
sightlines passing through discs/haloes of foreground galaxies.}, discovered through our ongoing efforts to characterize the cold atomic gas phase around 
low-$z$ galaxies \citep{gupta2010,gupta2013,srianand2013,dutta2015,srianand2015}. This QGP consists of a radio-loud quasar (QSO), \sys (J004126.01$-$014315.6, 
also known as UM~266; \zem = 1.679), at an impact parameter ($b$) $\sim$25~kpc from the spiral galaxy, \gal (J004121.53$-$014257.0; $z_g$ = 0.0177). This QGP 
is special in three regards. First, the background radio source shows multiple components spread over few arcseconds. \sys consists of a radio core component 
and two lobes spread over $\sim20''$ in the 1.4 GHz continuum map \citep{downes1986}, with the structures resolving into compact clumps in the higher frequency 
maps \citep[][5 GHz and 8.4 GHz respectively]{barthel1988,fernini2014}. High continuum flux of individual components provides us with the unique opportunity to 
use \hon\ \21 absorption to probe the distribution and structure of the CNM gas up to $\sim$7~kpc at the redshift of the foreground galaxy. Secondly, the radio 
source is located well outside the optical stellar disc of UGC~00439, allowing us to probe conditions in the extended $\hon$ disc. Finally, the foreground galaxy 
is close enough to detect \hon\ \21 emission and carry out a joint \hon\ \21 emission/absorption line analysis to address some of the above mentioned issues. 
Till date about 12 low-$z$ \hon\ \21 absorbers have been detected from the observations of QGPs \citep[see][in addition to the above mentioned studies of QGPs]{carilli1992, borthakur2010, borthakur2011, Reeves15, zwaan2015}.
Most of these are associated with the stellar discs of late-type galaxies and are towards quasars with radio emission compact at kpc scales, and joint analysis 
of absorption and emission measurements have been performed in very few cases \citep[see][]{stocke1991,stocke2010,keeney2005,keeney2011}.

This paper is structured as follows. In Section \ref{sec_ugc}, we summarize various properties of the host galaxy, UGC~00439, as provided in 
the literature. In Section \ref{sec_observations}, we describe our Giant Metrewave Radio Telescope (GMRT) observations of the system. Next, we 
give details of the GMRT $\hon$ \21 absorption towards the background radio source, of the GMRT $\hon$ \21 emission from the galaxy, and of the 
metal absorption lines detected in the optical spectrum of the QSO obtained using the Keck High Resolution Echelle Spectrometer (HIRES), in Section 
\ref{sec_results}. We then discuss the origin of the \hon\ \21 absorber in terms of the host galaxy properties in Section \ref{sec_discussion}. We 
conclude by summarizing our results in Section \ref{sec_summary}. Throughout this work we have adopted a flat cosmology with $H_{\rm 0}$ = 70\,\kms~Mpc$^{-1}$ and $\Omega_m$ = 0.27.
%
%
\section{Properties of UGC\,00439}  
\label{sec_ugc}  
The galaxy, UGC~00439, is part of the Near Field Galaxy Survey \citep{jansen2000a,jansen2000b}, which provides the U-, B-, R-band integrated 
and nuclear spectrophotometry. In B-band, \gal has an absolute magnitude, $M_B$ $\sim-$20.6, luminosity, log~$L_B$ $\sim$10.4 $L_\odot$, 
and radius at which the surface brightness drops to 25 mag arcsec$^{-2}$, $r_{25}^B$ $\sim$36$''$ (13\,kpc). From the optical photometry, 
\gal is classified as a face-on (inclination angle, $i$ = 0\textdegree) Sa galaxy having a position angle (PA) of the major axis (measured 
north through east) of 170\textdegree. However, from the Two Micron All-Sky Survey (2MASS) K\_s band photometry, \gal is estimated to have 
$i$ = 31\textdegree~and PA of major axis = 160\textdegree~\citep{skrutskie2006}. \gal is a star-forming (integrated SFR(H$\alpha$) 
$\sim$5 $M_\odot~yr^{-1}$) galaxy with solar metallicity (log(O$/$H) $+$ 12 $\sim$8.9) \citep[see for details][]{kewley2004}. \citet{kannappan2013} 
estimated the stellar mass of \gal as, log~$M_*$ $\sim$10.5 $M_\odot$. From the correlation found between the optical B-band magnitude 
and the $\hon$ size of low-$z$ galaxies \citep{broeils1997,lah2009}, the diameter within which half the $\hon$ mass of the galaxy is 
expected to be contained is $\sim$28~kpc. The Arecibo single dish $\hon$ \21 emission spectrum of the galaxy gives the velocity width 
of the emission as 115 $\pm$ 2\,\kms~and the total integrated flux density as, $\int S dv = 7.07 \pm 0.76$~Jy\,\kms~\citep{springob2005}. 
This total flux density translates into an $\hon$ mass, log~$M$($\hon$) = 9.99 $\pm$ 0.05 $M_\odot$. The $\hon$ \21, optical and ultraviolet 
(UV) observations in the literature of \gal satisfy the standard scaling relations obtained from the Arecibo Legacy Fast Arecibo L-band Feed Array 
($\alpha$.40), Sloan Digital Sky Survey (SDSS) and Galaxy Evolution Explorer (GALEX) surveys \citep{huang2012}. The optical spectra of 
the nuclear region of the galaxy from the Six-degree-Field Galaxy Survey \citep{jones2009} gives its redshift as $z$ = 0.017694 $\pm$ 0.000150 
(heliocentric velocity, $V_{\rm H}$ = 5304.5 $\pm$ 45\,\kms), which matches well with that obtained from the Arecibo $\hon$ \21 emission ($z$ = 0.017686 $\pm$ 0.000005 
and $V_{\rm H}$ = 5302 $\pm$ 2\,\kms).
%
%
\section{Radio observations}  
\label{sec_observations}  
The QSO \sys was observed twice using the GMRT on 2014 October 17 (2.6h on source) and 2015 July 4 (4.5h on source). The 
observing setup for the first run was optimized to detect \hon\ \21 absorption at the redshift of the foreground galaxy, 
UGC~00439, using a baseband bandwidth of 4.17 MHz split into 512 channels (velocity resolution $\sim$1.8\,\kms). Whereas 
in the second run, we used maximum available bandwidth to cover the $\hon$ \21 emission associated with UGC~00439, i.e. 
a baseband bandwidth of 16.67 MHz split into 512 channels (velocity resolution $\sim$7.1\,\kms). In both observations 
the pointing centre was at RA = 00$^{h}$41$^{m}$26.01$^{s}$, Dec = $-$01\textdegree43$'$15.59$''$ (J2000), and the band was 
centred on the redshifted \hon\ \21 frequency of UGC~00439. Standard calibrators were regularly observed during both the 
observations for flux density, bandpass, and phase calibrations. The data were reduced using the National Radio Astronomy 
Observatory (NRAO) Astronomical Image Processing System ({\tt AIPS}) following standard procedures as in \citet{gupta2010}. 
After initial flagging and gain calibration, a continuum map of the target source was made excluding channels at the edge and 
the centre of the band where the emission/absorption is expected. Using this map as a model, self-calibration complex gains 
were determined which were also applied to all the frequency channels. 
The self-calibrated continuum map was used to subtract the continuum emission from the visibility data set using the {\tt AIPS} 
task {\tt `UVSUB'}. Then the heliocentric correction was applied to the visibility data set using the {\tt AIPS} task {\tt `CVEL'}. 
The analyses of \hon\ \21 absorption towards \sys as described in Sections \ref{sec_radiocontinuum}, \ref{sec_21cmabsorption} and 
\ref{sec_21cmextent} is based on the data of 2014 October 17, and the analyses of \hon\ \21 emission from UGC~00439 as described in 
Section \ref{sec_21cmemission} is based on the data of 2015 July 4. The continuum subtracted visibility data sets from these were 
subjected to imaging with appropriate weightings to detect the \hon\ \21 absorption and emission. Details of the imaging parameters 
and properties of the images are described in the following sections.
%
%
\section{Analyses}
\label{sec_results}
\subsection{Radio continuum of \sys}
\label{sec_radiocontinuum}
\begin{figure}
\includegraphics[width=0.5\textwidth, bb=40 125 570 670, clip=true]{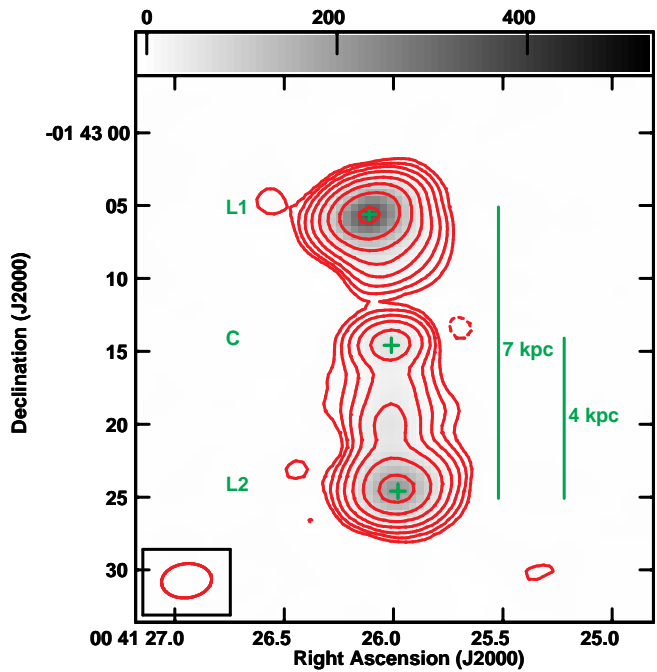}
\caption{GMRT image of \sys at 1.4 GHz. The contour levels are plotted as 2.5 $\times$ ($-$1,1,2,4,8,...)\,mJy~beam$^{-1}$. Note that solid 
lines correspond to positive values while dashed lines correspond to negative values. At the bottom left corner of the image the restoring 
beam is shown as an ellipse. The size and PA of the beam are $3.44''\times2.34''$ and $-$83.60\textdegree~respectively. The 
continuum peaks of the core, the northern lobe and the southern lobe are identified as C, L1 and L2 respectively. The vertical lines  
indicate the projected separation between these components at the redshift of UGC~00439.}
\label{fig_gmrtcontours}
\end{figure}

The radio source, J0041$-$0143, in our GMRT continuum image (see Fig.~\ref{fig_gmrtcontours}), consists of a core co-spatial with the 
optical QSO and two lobes separated by $\sim20''$. The continuum image, made using {\tt ROBUST=0} weighting and no {\it uv}-taper in 
the {\tt AIPS} task {\tt `IMAGR'}, is of spatial resolution $3.44''\times2.34''$, and has root mean square noise (rms) of 0.5\,\mjb 
near the target source and rms of 0.2\,\mjb away from it. The integrated flux density of the radio source in our image is 1026 mJy, 
consistent with that reported in the Faint Images of the Radio Sky at Twenty-Centimeters (FIRST; 1040 mJy) and the NRAO VLA Sky Survey 
(NVSS; 1034 mJy) catalogs. Note that the higher spatial resolution of the GMRT image, as compared to that of FIRST ($\sim5''$) and NVSS 
($\sim45''$), allows us to resolve all the three radio components distinctly. The radio `core', i.e. the component coincident with the 
optical QSO, and the radio continuum peaks associated with the northern and the southern lobes are labelled as C, and L1 and L2 respectively.
\subsection{$\hon$ \21 absorption towards \sys}
\label{sec_21cmabsorption}
The continuum-subtracted data set was imaged using the same weighting and beam as the {\tt ROBUST=0} continuum image described above (and 
shown in Fig.~\ref{fig_gmrtcontours}).  
Due to the lack of short baselines which were affected by radio frequency interference and flagged, and the choice of {\tt ROBUST=0} 
weighting, no $\hon$ \21 emission was detected in the cube. 
For deconvolution of $\hon$ \21 absorption signal while imaging the continuum-subtracted visibility data set in the {\tt AIPS} task 
{\tt `IMAGR'}, a mask true only for pixels $>$2.5\,\mjb in the radio continuum image (Fig.~\ref{fig_gmrtcontours}) was used.
The final image cube has rms $\sim$2.0\,\mjb~channel$^{-1}$. The spectra at the location of the radio components of interest were extracted 
from this cube, and if necessary, a first-order cubic spline was fitted to remove the residual continuum from the spectra. 

\begin{figure*}
\vbox{
\includegraphics[width=0.3\textwidth, angle=90]{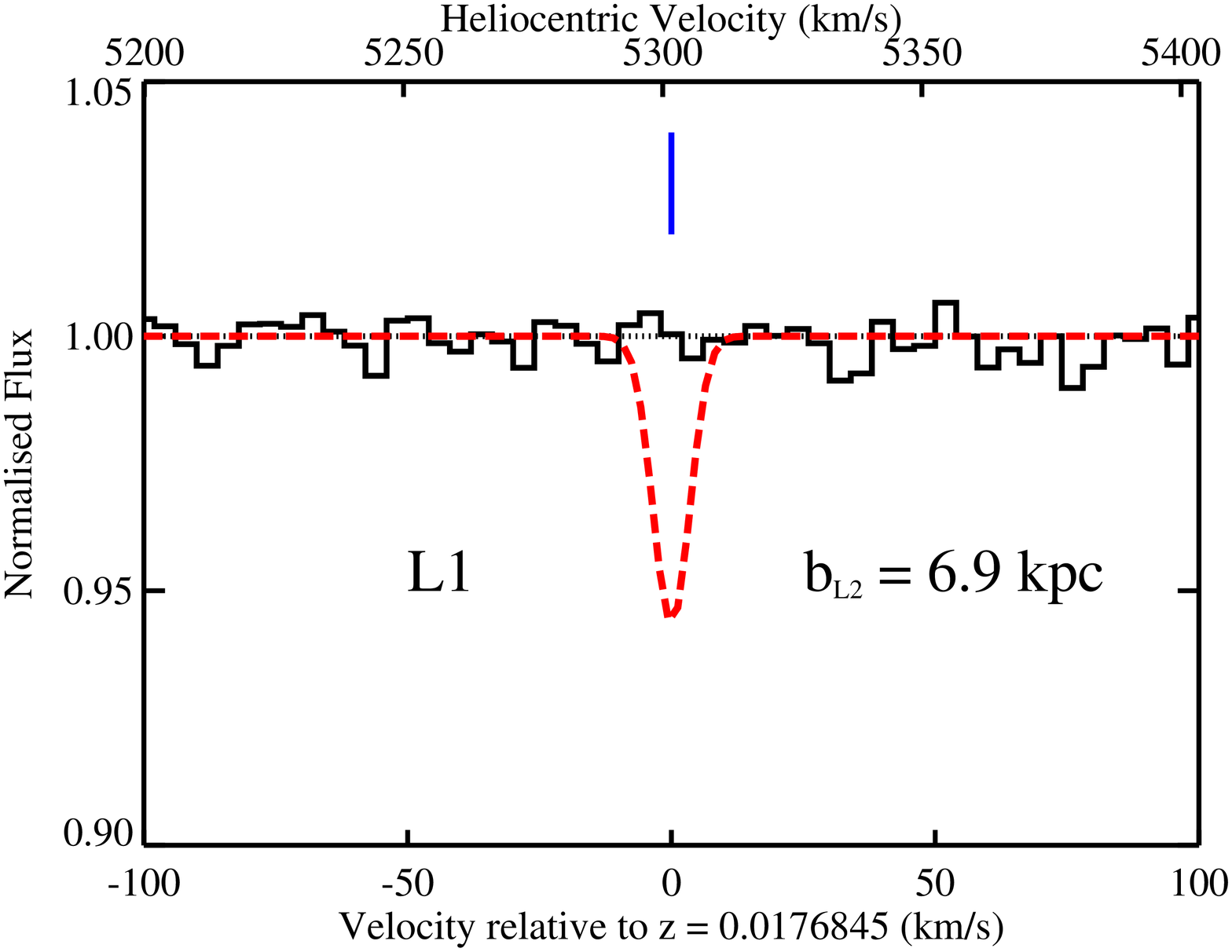}
\includegraphics[width=0.3\textwidth, angle=90]{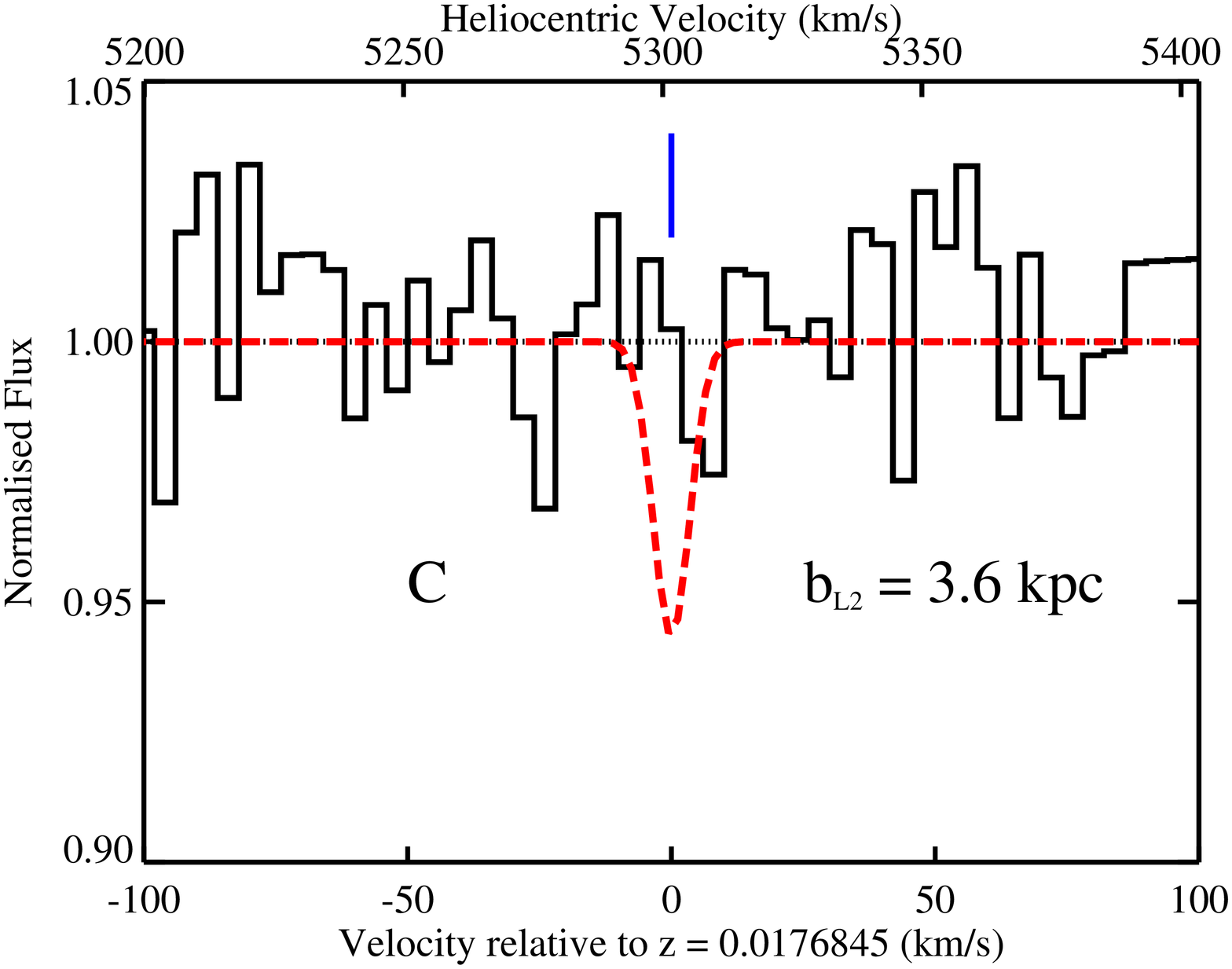}
\includegraphics[width=0.3\textwidth, angle=90]{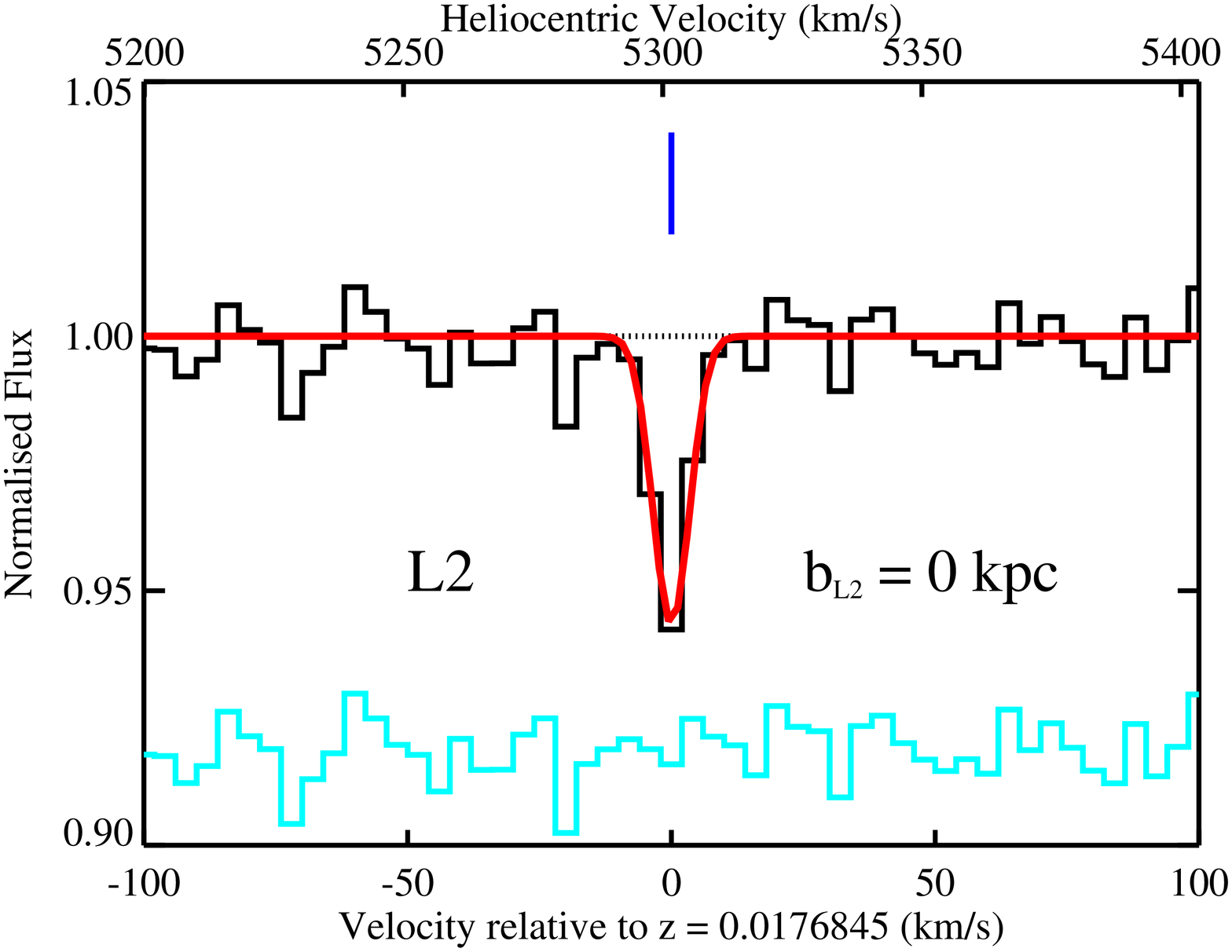}
}
\caption{GMRT \hon\ \21 absorption spectra towards the continuum components L1 {\it (Top Left)}, C {\it (Top Right)} and L2 {\it (Bottom)}, 
as marked in Fig.~\ref{fig_gmrtcontours}. The spectra are smoothed to 4\,\kms~for display purpose. In the bottom panel the best-fitting 
single Gaussian profile to the \hon\ \21 absorption towards L2 is overplotted in solid red line, and the residuals from the fit 
(shifted in the flux scale for display) are plotted below in cyan. This fit is also overplotted in the other two panels in dashed red 
line for comparison. The vertical tick marks the position of the peak optical depth detected towards L2. $b_{L2}$ given in each panel 
is the projected separation of the background continuum emission with respect to L2 at the redshift of UGC~00439, as also indicated in Fig.~\ref{fig_gmrtcontours}.}
\label{fig_21cmspec}
\end{figure*}

We detect \hon\ \21 absorption towards L2 at the systematic redshift of UGC~00439, with a maximum line depth of 16 mJy (i.e. at 8$\sigma$ 
significance). The \hon\ \21 absorption towards L2 is spread over 10 channels or $\sim$18\,\kms, with peak optical depth, $\tau_p$ = 0.08 $\pm$ 0.01, 
and total integrated optical depth, \taudv = 0.52 $\pm$ 0.07\,\kms. However, we do not detect any \hon\ \21 absorption towards L1 or C. The 
optical depth towards L1 and C has to be less than that towards L2 by a factor of $\gtrsim$7 and of $\gtrsim$2 respectively (at 3$\sigma$ level). 
The \hon\ \21 absorption spectra towards L1, C and L2 are shown in Fig.~\ref{fig_21cmspec}, while the parameters derived from these spectra are 
provided in Table~\ref{tab:radioparameters}. 

The \hon\ \21 absorption profile towards L2 shows a clear evidence of arising from the CNM phase. Although the CNM phase is observed to exhibit a 
range of temperatures ($\sim$20$-$200~K), for the sake of simplicity we adopt throughout this paper a typical CNM mean temperature of 100~K, as 
observed in the Milky Way \citep{heiles2003}, to infer the physical nature of the \hon\ \21 absorption component. 
In the absence of subarcsecond-scale spectroscopic observations, we assume unit covering factor ($f_c$) of the cold \hon\ absorbing gas within 
the beam. The $\hon$ \21 optical depth towards L2 translates to a column density of $\hon$ gas in the CNM phase of 
9.5 $\times$ 10$^{19}$ $\times$ ($T_s/100$~K)($1/f_c$)\,\cms\ (see Table \ref{tab:radioparameters}). The \hon\ \21 optical depth 
constraints towards L1 and C lead to $N$($\hon$) $<$ 1.3 $\times$ 10$^{19}$ $\times$ ($T_s/100$~K)($1/f_c$)\,\cms\ and 
$N$($\hon$) $<$ 5.1 $\times$ 10$^{19}$ $\times$ ($T_s/100$~K)($1/f_c$)\,\cms\ in the CNM phase respectively.

A single Gaussian is adequate to fit the \hon\ \21 absorption line detected towards L2. The best-fitting Gaussian to the \hon\ \21 absorption line 
(as shown in Fig.~\ref{fig_21cmspec}) has a full width at half-maximum, FWHM = 8 $\pm$ 1\,\kms. This constrains the kinetic temperature as $\le$1400~K.  
In nearby spiral galaxies, CNM and WNM phases have been identified through decomposition of \hon\ \21 emission lines into narrow (average dispersion, 
$\sigma\sim$7\,\kms) and broad ($\sigma\sim$17\,\kms) components \citep[][]{Tamburro2009}. The average dispersion of the narrow component observed 
in these galaxies at $r\sim$2$r_{25}$ (i.e. similar to the J0041$-$0043 - UGC\,00439 impact parameter) is $\sim$6\,\kms\ (FWHM $\sim$14\,\kms) 
implying $T\le$4300~K. Thus, it appears that in the outer discs of galaxies, the thermal broadening is much less effective than the turbulence 
in the gas. We come back to this point in Section~\ref{sec_21cmextent}.
\begin{table*} 
\caption{Details of \hon\ \21 absorption towards different radio sightlines of J0041$-$0143.}
\centering
\begin{tabular}{cccccccccc}
\hline
\hline
Sightline & Coordinates & $A_{L2}$ & $b_{L2}$ & Peak Flux    & Spectral rms    & $\tau_p$ & $\int\tau dv$ & log~$N$($\hon$)        & $v_{L2}$ \\
          &   (J2000)   & ($''$)   & (kpc)    & (mJy~        & (\mjb           &          & (\kms)        & ($T_s/100$~K)($1/f_c$) & (\kms)   \\
          &             &          &          & beam$^{-1}$) & channel$^{-1}$) &          &               & (\cms)                 &          \\
(1)       & (2)         & (3)      & (4)      & (5)          & (6)             & (7)      & (8)           & (9)                    & (10)     \\
\hline
\multicolumn{10}{c}{\it Continuum peaks of the core and the lobes} \\
\hline
L1        & 00:41:26.11 $-$01:43:05.59 & 19.1 & 6.9 & 354.0 & 2.3 & $<$0.01         & $<$0.07          & $<$19.10        & ---              \\
C         & 00:41:26.01 $-$01:43:14.59 & 10.0 & 3.6 &  66.7 & 2.1 & $<$0.03         & $<$0.28          & $<$19.71        & ---              \\
L2        & 00:41:25.98 $-$01:43:24.59 &    0 &   0 & 215.1 & 2.1 & 0.08 $\pm$ 0.01 & 0.52 $\pm$ 0.07 & 19.98 $\pm$ 0.05 & 0                \\
\hline
\multicolumn{10}{c}{\it Locations in the southern lobe} \\
\hline
1         & 00:41:25.98 $-$01:43:22.59 &  2.0 & 0.7 &  96.3 & 2.5 & 0.09 $\pm$ 0.02 & 0.35 $\pm$ 0.16 & 19.80 $\pm$ 0.20 & $-$3.6 $\pm$ 2.3 \\
2         & 00:41:26.18 $-$01:43:24.09 &  3.0 & 1.1 &  37.9 & 2.3 & 0.19 $\pm$ 0.07 & 1.20 $\pm$ 0.42 & 20.34 $\pm$ 0.15 & $+$1.1 $\pm$ 1.2 \\
3         & 00:41:25.78 $-$01:43:25.09 &  3.0 & 1.1 &  28.2 & 1.9 & 0.12 $\pm$ 0.06 & 1.14 $\pm$ 0.48 & 20.32 $\pm$ 0.18 & $+$4.6 $\pm$ 1.3 \\
4         & 00:41:26.01 $-$01:43:20.59 &  4.0 & 1.5 &  45.6 & 2.1 & 0.13 $\pm$ 0.05 & 0.65 $\pm$ 0.33 & 20.07 $\pm$ 0.22 & $+$3.9 $\pm$ 1.7 \\
\hline
\hline
\end{tabular}
\label{tab:radioparameters}
\begin{flushleft}
Column 1: Radio sightline (as labelled in Figs.~\ref{fig_gmrtcontours} and \ref{fig_southlobe}). 
Column 2: J2000 coordinates of the radio sightline.
Columns 3 \& 4: Angular (in arcsec) and projected (in kpc) separation at the redshift of UGC~00439 with respect to L2, respectively.
Column 5: Peak continuum flux in mJy~beam$^{-1}$.
Column 6: Spectral rms in mJy~beam$^{-1}$~channel$^{-1}$ at velocity resolution of 1.8\,\kms.
Column 7: Peak \hon\ \21 optical depth or $1\sigma$ limit to it in case of non-detections.
Column 8: Integrated \hon\ \21 optical depth or $3\sigma$ upper limit in case of non-detections with data smoothed to 10 km~s$^{-1}$.
Column 9: log~$N$($\hon$) (in \cms) corresponding to the optical depth given in column 8 for $T_s = 100$~K and $f_c = 1$.
Column 10: Optical depth weighted mean velocity (in \kms), in case of detections, with respect to the velocity of peak optical depth detected towards L2.  
\end{flushleft}
\end{table*}
\subsection{Extent and structure of \hon\ \21 absorbing gas}
\label{sec_21cmextent}
\begin{figure*}
\vbox{
\includegraphics[width=0.33\textwidth, trim=3cm 11cm 3cm 2cm]{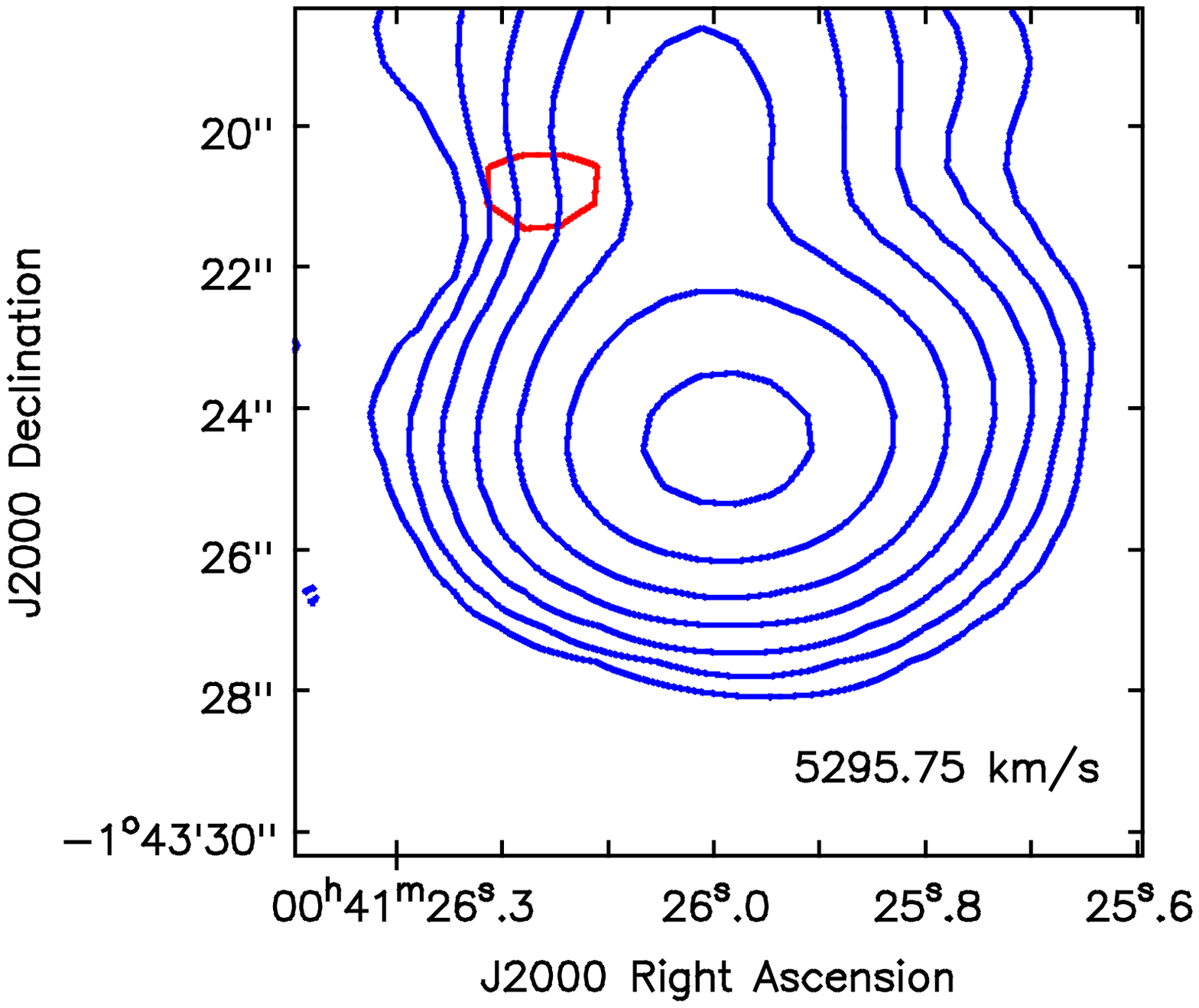}
\includegraphics[width=0.33\textwidth, trim=3cm 11cm 3cm 2cm]{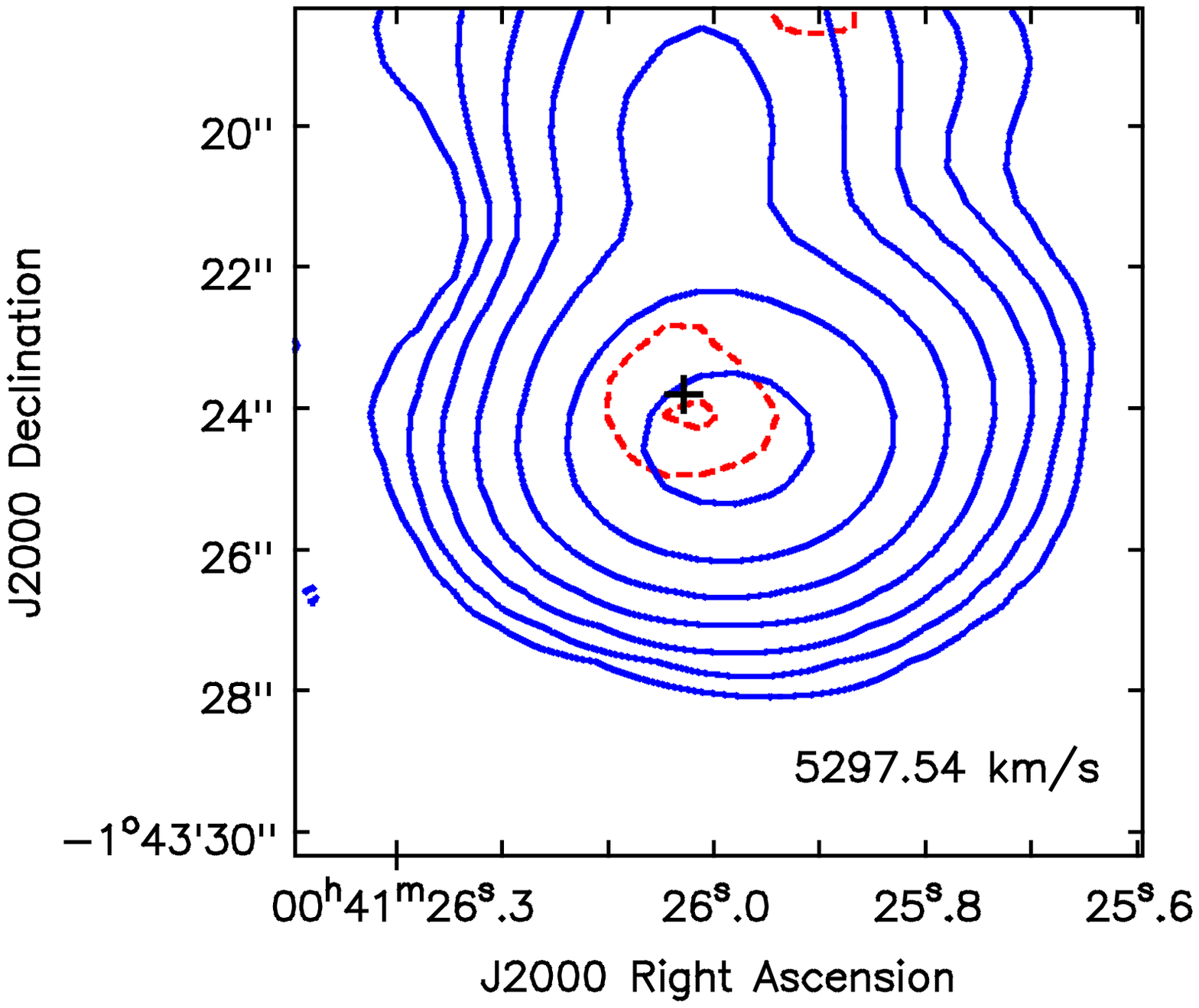}
\includegraphics[width=0.33\textwidth, trim=3cm 11cm 3cm 2cm]{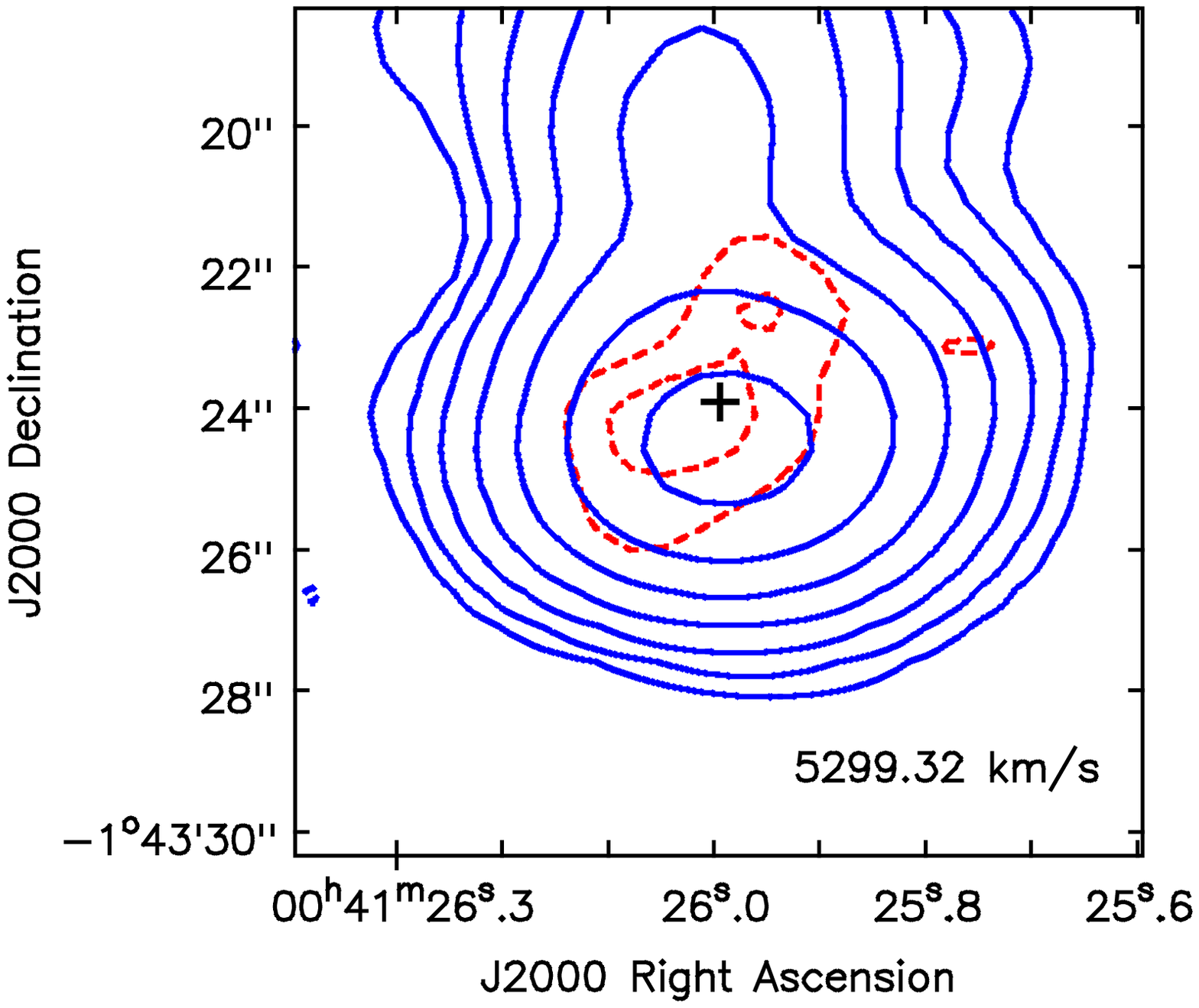}
\includegraphics[width=0.33\textwidth, trim=3cm 11cm 3cm 2cm]{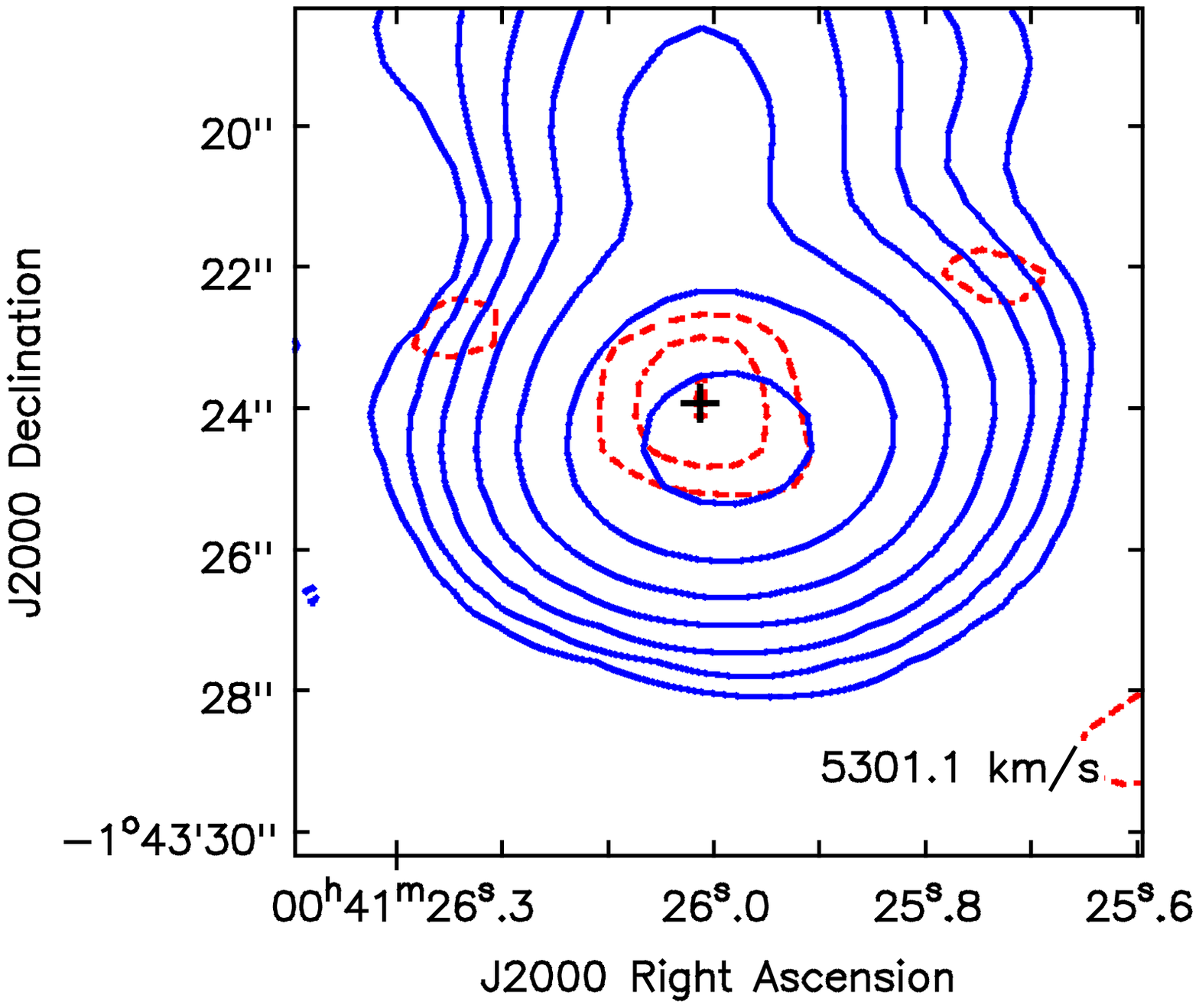}
\includegraphics[width=0.33\textwidth, trim=3cm 11cm 3cm 2cm]{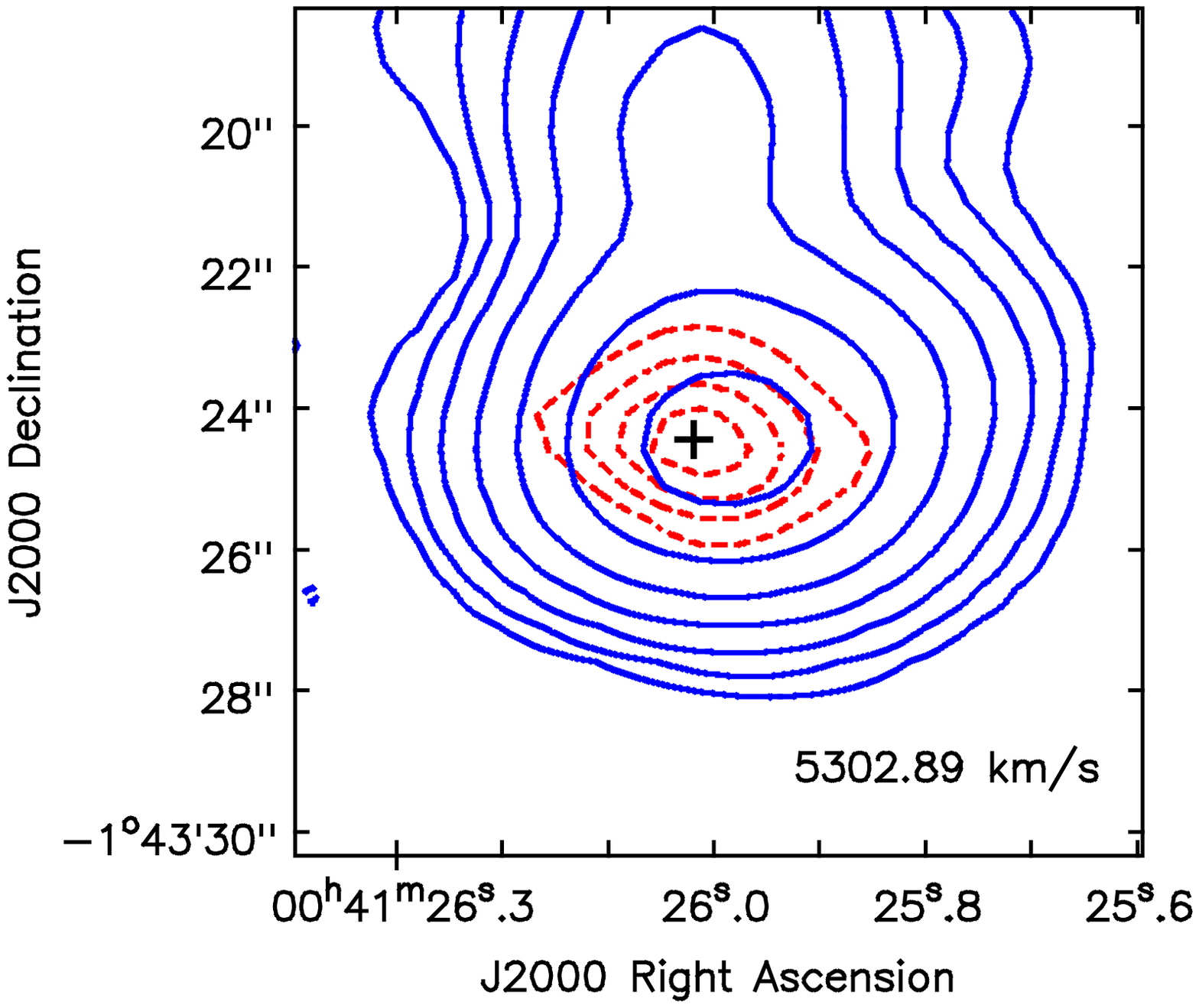}
\includegraphics[width=0.33\textwidth, trim=3cm 11cm 3cm 2cm]{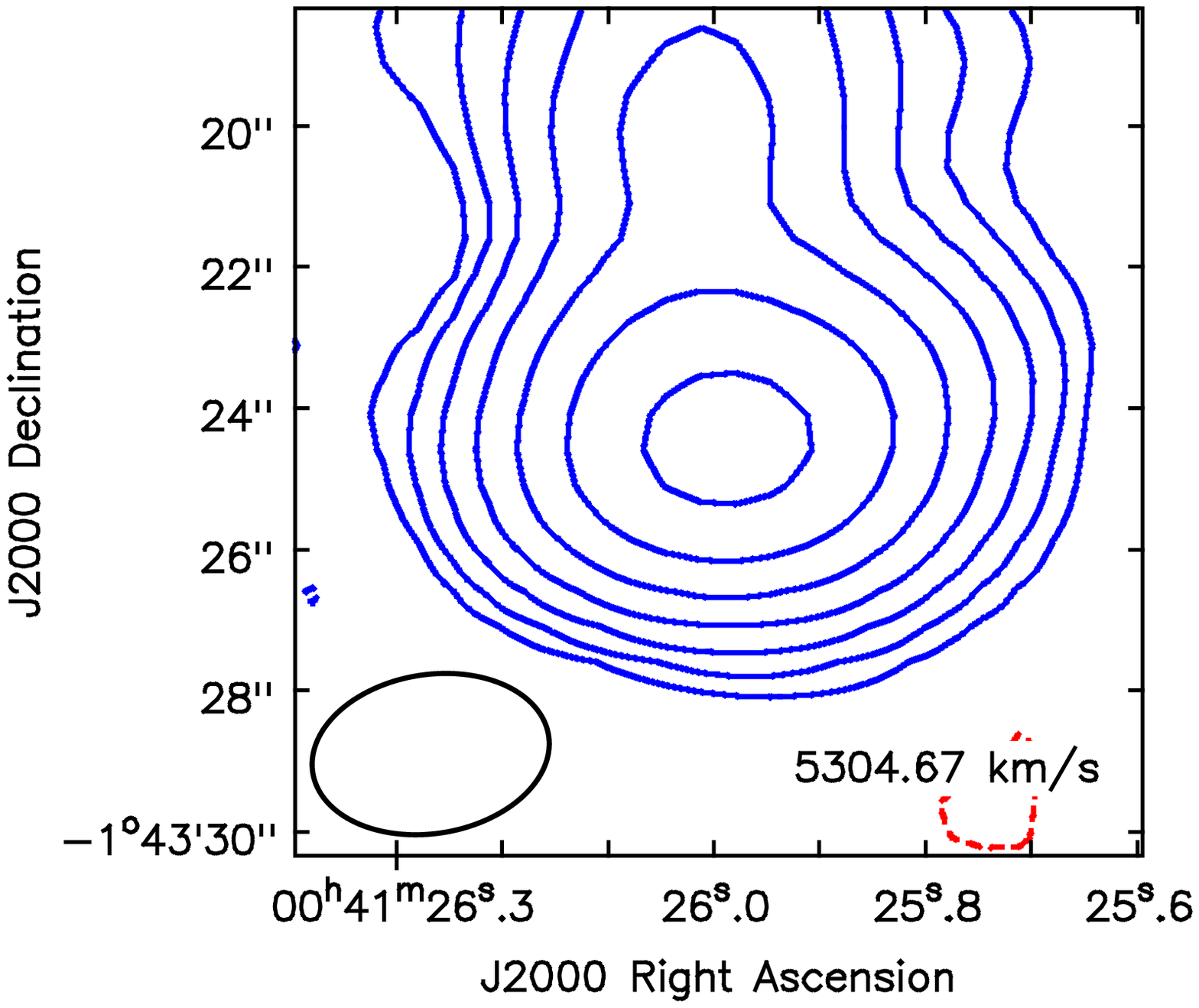}
}
\caption{Channels maps of the \hon\ \21 absorption towards \sys separated by 1.8\,\kms. The blue contours and the beam are the same as in 
Fig.~\ref{fig_gmrtcontours}. The red contours show the \hon\ \21 absorption with the levels plotted as 6.0 $\times$ ($-$2.5, $-$2.0, $-$1.5, 
$-$1.0, 1.0, 1.5, 2.0, 2.5) \mjb. The crosses mark the centre of the absorption obtained from two-dimensional single Gaussian fit. Note that 
solid lines correspond to positive values while dashed lines correspond to negative values. Heliocentric velocity of each channel is also given.}
\label{fig_absmaps}
\end{figure*}

\begin{figure*}
\vbox{
\includegraphics[width=0.48\textwidth, bb=40 190 580 600, clip=true]{J0041_MAP_SLOBE.PS}
\includegraphics[width=0.38\textwidth, angle=90]{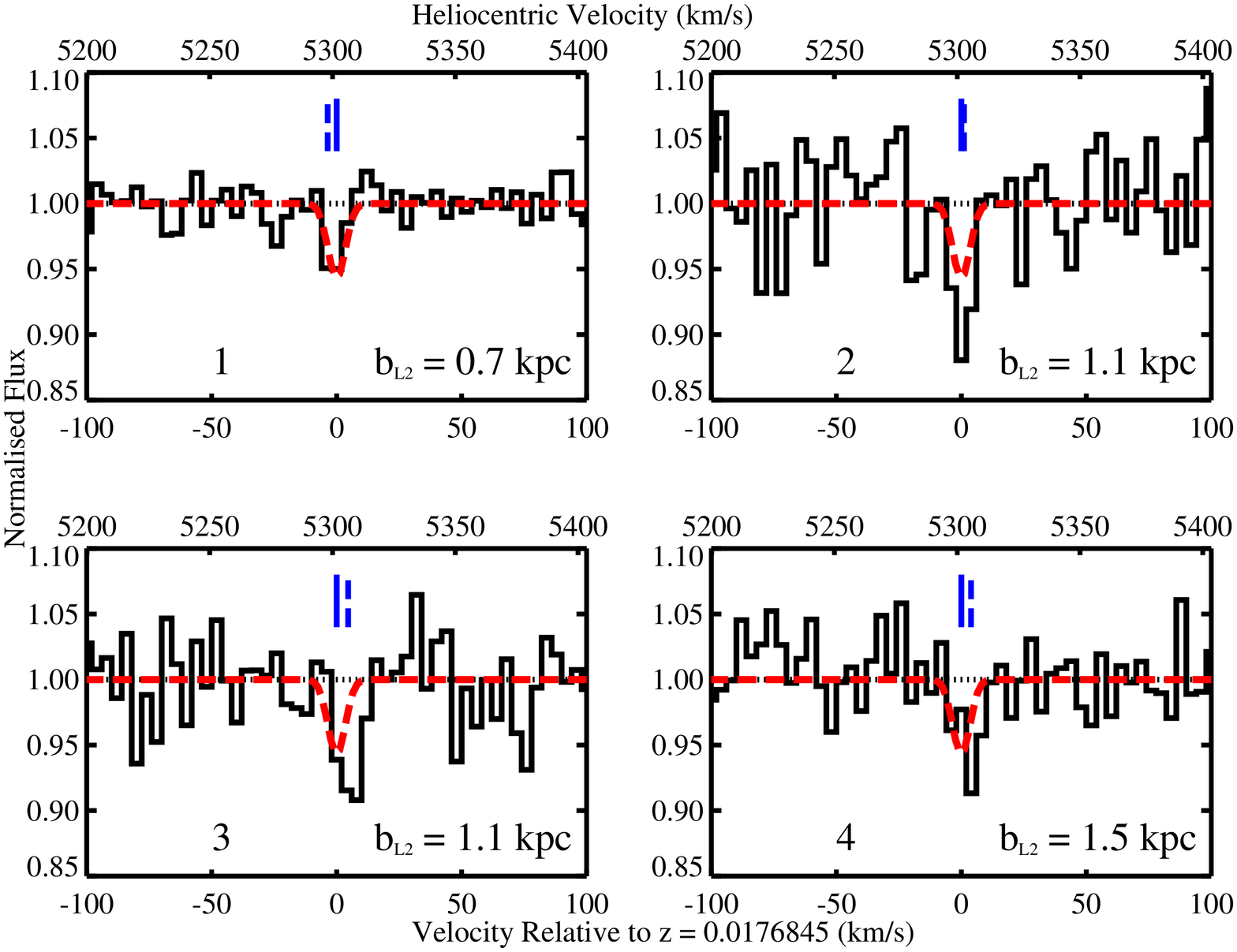}
}
\caption{{\it Left:} The contours and the beam are the same as in Fig.~\ref{fig_gmrtcontours} with the southern lobe zoomed in. Marked are 
the different sightlines, separated by the beam size, towards which \hon\ \21 absorption spectra are extracted. {\it Right:} The GMRT \hon\ 
\21 absorption spectra (smoothed to 4\,\kms~for display purpose) towards the sightlines 1, 2, 3 and 4 as marked in the figure to the left. Overplotted 
in dashed red line is the best-fitting Gaussian profile to the \hon\ \21 absorption towards L2, as shown in the bottom panel of Fig.~\ref{fig_21cmspec}, 
for comparison. The vertical solid tick marks the position of the peak optical depth detected towards L2, while the dashed tick marks the position 
of the optical depth weighted mean velocity of the absorption. $b_{L2}$ is as defined in Fig.~\ref{fig_21cmspec}. Details of the spectra extracted
from different locations can be found in Table \ref{tab:radioparameters}.}
\label{fig_southlobe}
\end{figure*}

The extended nature of the background radio source allows us to study the spatial extent and structure of the gas giving rise to the \hon\ \21 
absorption. As indicated above, the absorbing gas may extend up to the core ($\sim$3.6~kpc from L2) of J0041$-$0143, with optical depth dropping 
by a factor of $\gtrsim$2 between the lines-of-sight towards L2 and C. However, it is unlikely to extend up to the northern lobe ($\sim$7~kpc
from L2) since the optical depth towards L1 is at least factor of 7 less than that towards L2. In Fig.~\ref{fig_absmaps}, we show the channel 
maps of the \hon\ \21 absorption detected towards the southern lobe (dashed contours), overlaid on the continuum contours of \sys for reference. 
The \hon\ \21 absorption is detected at a significance of $\ge$3$\sigma$ over 4 channels or 7.2\,\kms. From Fig.~\ref{fig_absmaps}, we can see that 
the \hon\ \21 absorption has an extended structure, and is spread over a region larger than that of the beam (i.e. $\gtrsim$ 1~kpc$^{2}$) in two 
channels (having $V_{\rm H}$ = 5299\,\kms~and 5303\,\kms~respectively). To check the spatial extent of the absorbing gas we extracted spectra towards different 
sightlines separated by the beam size in the southern lobe. In Fig.~\ref{fig_southlobe} we show the \hon\ \21 absorption spectra towards four such sightlines 
where \hon\ \21 absorption is detected at $\ge$2$\sigma$, and their details are provided in Table \ref{tab:radioparameters}. The spatially resolved 
extraction seems to indicate that the absorption is likely to extend up to $\sim$2~kpc$^{2}$, covering most of the southern lobe. Note that we have 
checked that all the absorption features are present in both the XX and YY polarization spectra for consistency. However, we caution that the optical 
depth sensitivity falls off rapidly away from L2, and the \hon\ \21 absorption towards these sightlines are at lower levels of significance ($\sim$2-3$\sigma$). 

In addition, from Fig. \ref{fig_absmaps}, we note that the location of the maximum absorption or optical depth does not always coincide with that of the 
peak radio continuum flux. This can be caused by the radio emission in the southern lobe resolving into structures at sub-kpc-scales, as indicated by the 5 GHz 
\citep{barthel1988} and the 8.4 GHz \citep{fernini2014} continuum maps of this source. However, it is interesting to note from this figure that the location of 
the maximum absorption shifts with velocity. From two-dimensional single component Gaussian fitting to the absorption, we find that the centre of the absorption 
shifts by 1$''$ over 5.4\,\kms. This seems to indicate that the absorbing gas itself has structures at sub-kpc-scales and a line-of-sight velocity dispersion of the 
order of few \kms. This is further reflected in the variation (by factor $\sim$3) seen in the optical depth of \hon\ \21 absorption towards different sightlines 
across the southern lobe (see Table \ref{tab:radioparameters}). Again from the same table, we see that, while the gas shows no systematic velocity gradient over the 
region of the southern lobe, it does seem to have small-scale (few \kms) turbulent motion. We argue that it is this non-thermal motion that dominates the velocity 
spread (FWHM $\sim$8\,\kms) of the absorption detected towards L2. In summary, the absorbing gas exhibits both large- (kpc) and small-scale (sub-kpc) structures, along 
with small-scale turbulent motion. Detailed understanding of the small-scale structures in the absorbing gas would greatly benefit from deeper \hon\ \21 observations.   
\subsection{Metal absorption towards \sys}
\label{sec_metalabsorption}
\begin{figure}
\includegraphics[width=0.4\textwidth, angle=90]{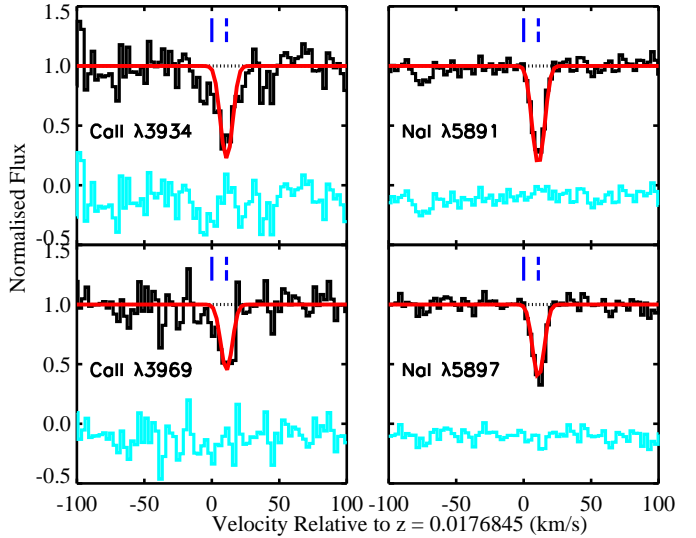}
\caption{The $\cat$ and $\nao$ absorption towards the QSO \sys detected in the Keck-HIRES spectrum. The optical emission of the QSO coincides
with component C. Best-fitting Voigt profiles are overplotted in red. The residuals from the fit are shown at the bottom in cyan. The vertical 
solid and dashed ticks mark the positions of the peak \hon\ \21 optical depth detected towards L2 and of the peak metal absorption respectively.}
\label{fig_metalspec}
\end{figure}

$\cat$ and $\nao$ absorption are detected towards the optical QSO \sys at $z$ = 0.017721 in the Keck HIRES spectrum from the 
Keck Observatory Database of Ionized Absorbers toward Quasars \citep[KODIAQ;][]{omeara2015}. We fit Voigt profiles to the 
metal lines using \textsc{vpfit}\footnote{http://www.ast.cam.ac.uk/$\sim$rfc/vpfit.html}. The best-fitting single component 
Voigt profile with Doppler parameter of 4.1 $\pm$ 0.3\,\kms~(or FWHM of 6.7 $\pm$ 0.5\,\kms) gives log~$N$($\cat$) = 12.40 $\pm$ 0.06 
and log~$N$($\nao$) = 12.29 $\pm$ 0.03. In Fig.~\ref{fig_metalspec}, we show the metal absorption and the best-fitting Voigt 
profiles. The metal absorption towards the optical QSO, which is cospatial with the radio core, indicates the presence of 
$\hon$ gas at this location. Moreover, since $\cat$ and $\nao$ are usually believed to be tracers of cold gas, the $\hon$ 
gas detected in absorption towards L2 is likely to extend up to the core (with factor $\gtrsim$2 variation in optical depth). 
Note that the peak of the metal absorption is redshifted by $\sim$11\,\kms~with respect to the peak \hon\ \21 optical depth 
towards L2. In other words, since the \hon\ \21 absorption occurs at the systematic redshift of UGC~00439, the metal 
absorption is redshifted with respect to the systematic redshift of the host galaxy.

In the Galactic ISM, $N$($\nao$)/$N$($\cat$) $>$ 1 for the observed values of $N$($\nao$) \citep[][Fig.~9]{welty1996}, 
while the observed ratio is $\sim1$ in the present case. This could mean that Ca depletion is not as high as that typically 
seen ($\sim$ $-$3 dex) in the Galactic ISM, and accordingly the $N$($\hon$) may not be very high since depletion usually 
correlates with it. If $N$($\nao$) scales with \nhi as seen in our Galaxy \citep{ferlet1985,wakker2000}, then we expect 
log~$N$($\hon$) $\sim$20.6. In that case, from the \hon\ \21 optical depth limit towards the core, the gas would have 
$T_s\gtrsim800$~K (for $f_c$ = 1). On the other hand, if we assume the gas to be cold ($\sim$100~K), we expect log~$N$($\hon$) 
$<$19.7 (for $f_c$ = 1; see Table~\ref{tab:radioparameters}), or $\gtrsim8$ times more $\nao$ per $\hon$ to be present 
in this gas than what is typically seen in the Galactic ISM. As $\nao$ is not the dominant ion of Na in the $\hon$ phase, 
this can be explained by the ionizing field at the location of the absorber being weaker than the mean Galactic radiation 
field, as expected in the outer extended $\hon$ disc. Hence, $\nao$ may not be a good tracer of $\hon$ in sightlines that 
pass away from the stellar disc of galaxies \citep[e.g.][]{dutta2015}.  
\subsection{$\hon$ \21 emission from \gal}
\label{sec_21cmemission}
\begin{figure}
\includegraphics[width=0.38\textwidth, angle=90]{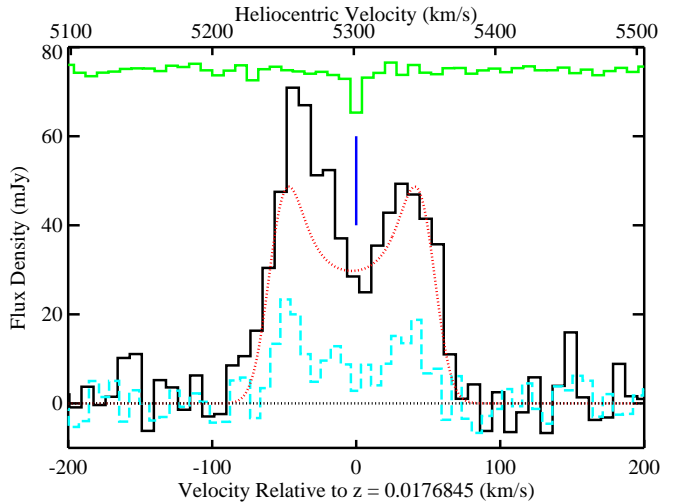}
\caption{The Arecibo $\hon$ \21 emission spectrum \citep{springob2005} in solid histogram and the GMRT $\hon$ \21 emission spectrum in dashed histogram 
of the galaxy, UGC~00439. The $\hon$ \21 emission spectrum expected from a simple rotating disc galaxy model (see Section \ref{sec_kinematics} 
for details) is overplotted in dotted line. Shown in green is the $\hon$ \21 absorption spectrum towards L2, smoothed to the velocity resolution 
of the Arecibo spectrum ($\sim$8\,\kms), and offset by 75~mJy along the y-axis for ease of comparison. The vertical tick marks the location of the 
peak optical depth found in the $\hon$ \21 absorption spectrum.}
\label{fig_hiemission}
\end{figure}

In order to detect $\hon$ \21 emission from the galaxy using GMRT data, we imaged the continuum subtracted visibilities with different 
{\it uv}-tapers and robust weightings to make spectral cubes of different spatial resolutions. We find that the $\hon$ \21 emission can 
be best measured using the present data at a spatial resolution of $46.4'' \times 33.4''$ (i.e. 17\,kpc $\times$ 12\,kpc). A coarser 
spatial resolution does not lead to increase in the total flux density detected from the galaxy, and the present data do not have 
sufficient signal-to-noise ratio to detect $\hon$ \21 emission reliably at a finer spatial resolution. Hence, the discussion on $\hon$ 
\21 emission presented in this paper is based on the image cube of spatial resolution $46.4'' \times 33.4''$ and velocity resolution 
$\sim$7.1\,\kms, made using {\tt ROBUST=1} weighting (rms $\sim$2 \mjb~channel$^{-1}$). 
While imaging, the $\hon$ \21 emission signal was deconvolved using mask(s) true only for pixels with absolute flux $>$ 3 times the single channel rms.
The global $\hon$ \21 emission spectrum is obtained from this cube by integrating the flux density over all the regions where $\hon$ \21 
emission associated with UGC~00439 is detected. The total integrated flux density of the $\hon$ \21 emission is, $\int S dv = 1.42 \pm 0.14$~Jy\,\kms. 
The GMRT $\hon$ \21 emission is detected over almost the full velocity range of the Arecibo emission, with the double peaks of the emission 
profiles matching in velocity (see Fig.~\ref{fig_hiemission} for comparison). 
Note that the total GMRT $\hon$ \21 emission recovers only $\sim$20\% of that found in the Arecibo spectrum. However, the flux recovered 
in the wings and the two peaks is $\sim$33\% of that in the Arecibo profile, i.e. there is no uniform offset in amplitude between the two profiles. 
This implies that the \hon\ \21 emission that is not detected by the GMRT data could be arising from a separate component of the $\hon$ gas 
associated with UGC~00439. The poor sensitivity and {\it uv}-coverage of the GMRT data perhaps does not allow us to detect any diffuse 
and extended \hon\ gas around UGC~00439. We discuss the structure and kinematics of the \hon\ gas associated with \gal further in Section~\ref{sec_kinematics}.

\begin{figure*}
\vbox{
\includegraphics[height=0.33\textheight, angle=270, bb=570 160 45 680, clip=true]{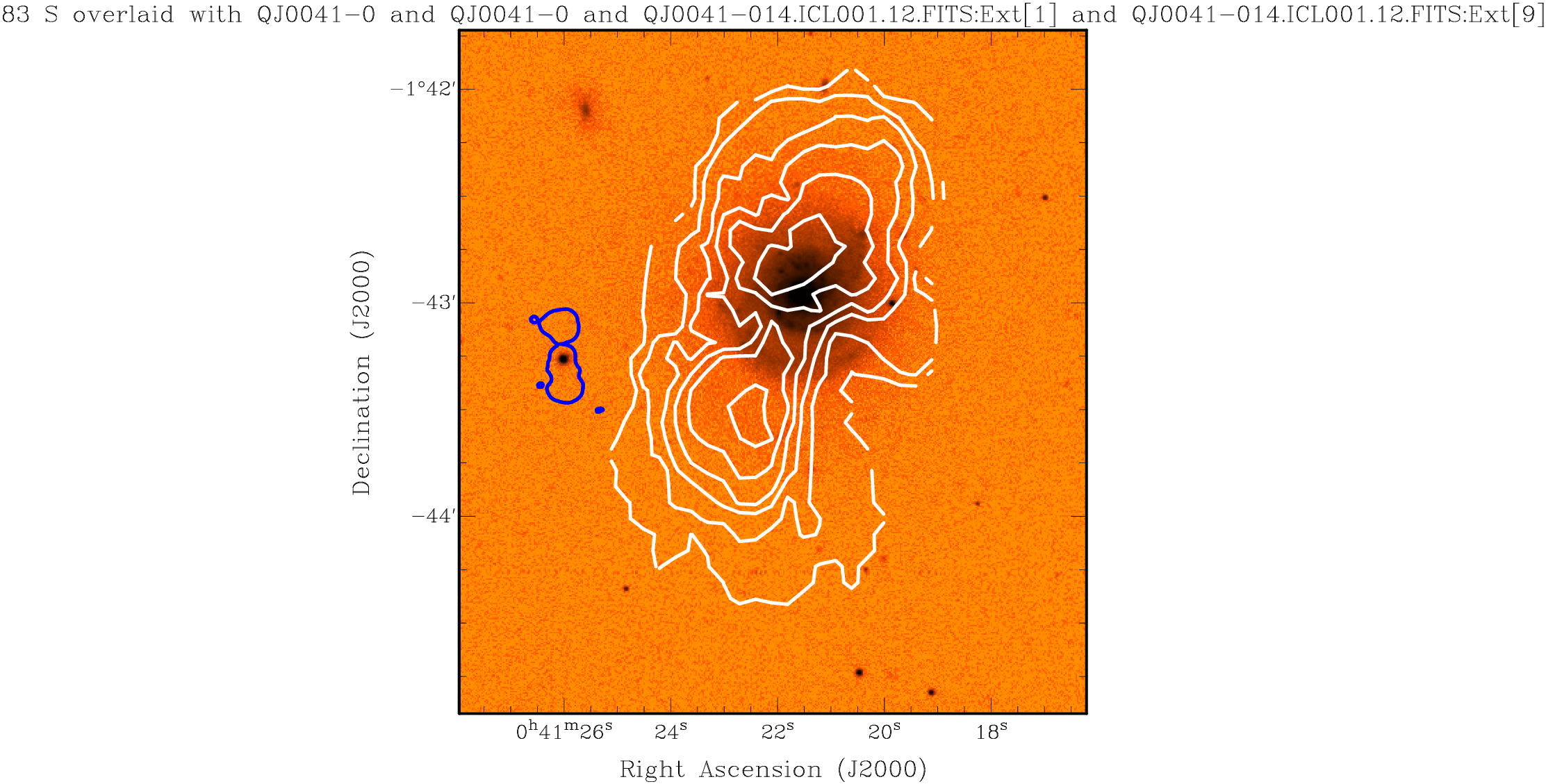}
\includegraphics[height=0.33\textheight, bb=65 188 560 608, clip=true]{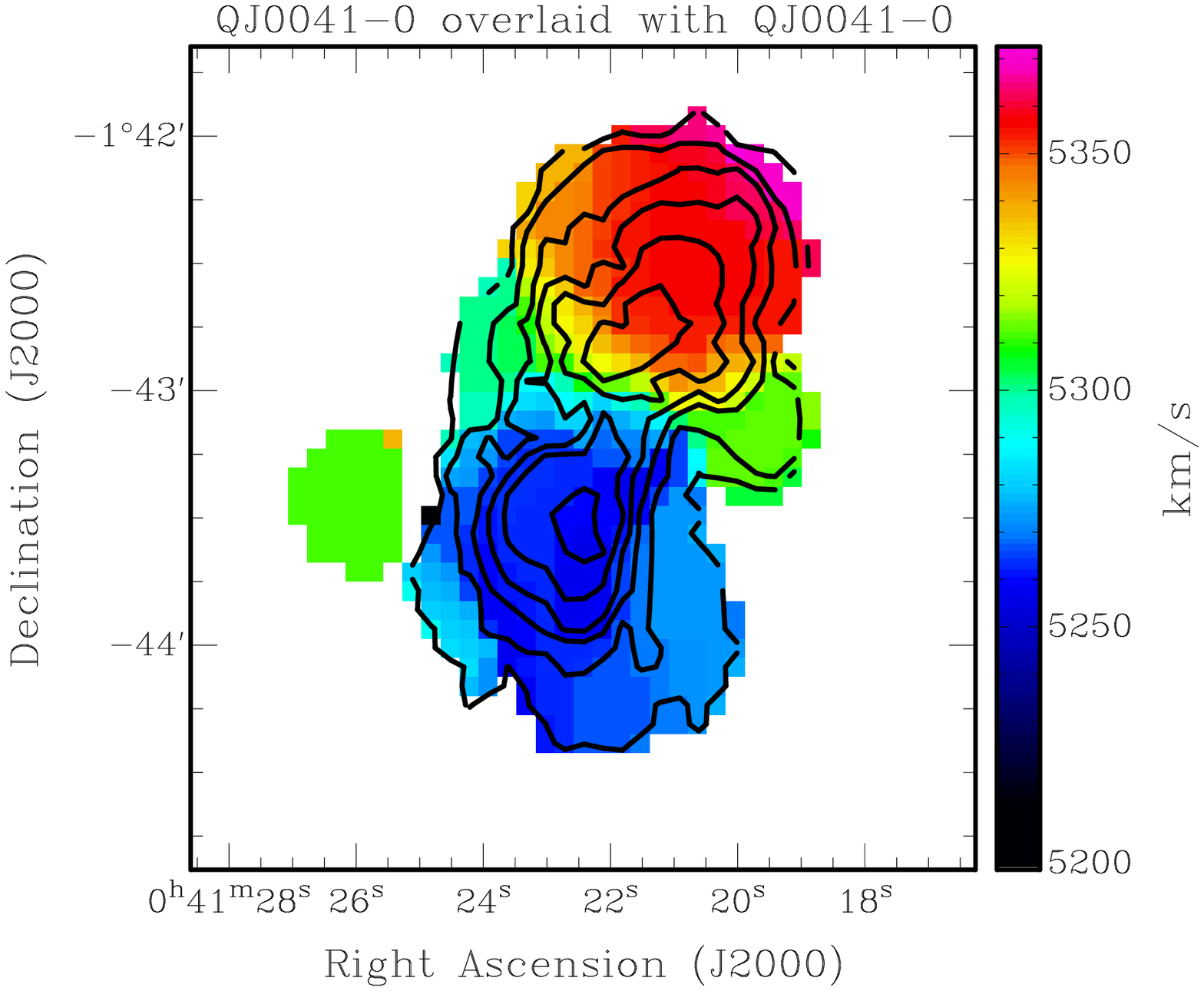}
}
\caption{{\it Left:} SDSS r-band image of the QGP, J0041$-$0143$/$UGC~00439, is shown. The GMRT $\hon$ \21 moment-0 map 
(of spatial resolution $46.4'' \times 33.4''$) is shown as a contour plot, with the levels corresponding to $N$($\hon$) 
= (0.4, 0.8, 1.2, 1.6, 2.0, 2.4) $\times$ 10$^{20}$\,\cms. The outermost contour of the GMRT radio continuum of \sys 
from Fig.~\ref{fig_gmrtcontours} (2.5\,mJy~beam$^{-1}$) is also shown for reference in blue. {\it Right:} The GMRT 
moment-1 map of \gal overlaid with contours of the moment-0 map as shown to the left. The green patch at the location 
of the quasar corresponds to the redshift of the $\hon$ \21 absorption. Note that from Arecibo $\hon$ \21 measurement, 
\gal has $V_H$ = 5302 $\pm$ 2\,\kms.}
\label{fig_mom0overlay}
\end{figure*}

For generating \hon\ \21 moments maps, we made a mask based on emission detected above a threshold of 3$\sigma$ noise in the original image 
cube and the cube smoothed over twice the spatial and the velocity resolution. 
We show the contours of the total \hon\ \21 (moment-0) map, corrected for the primary beam attenuation, overlaid on the SDSS r-band image 
in the left-hand panel of Fig.~\ref{fig_mom0overlay}, with the outermost level corresponding to log~$N$($\hon$) = 19.6, and the highest density clumps 
corresponding to log~$N$($\hon$) = 20.4. We can see that the $\hon$ \21 emission, detected in the GMRT data, extends beyond the optical disc of 
the galaxy, but does not extend up to the QSO, J0041$-$0143. We also checked in the channel maps of the cube which contain the \hon\ \21 absorption, 
that no $\hon$ \21 emission extends up to the location of the absorber. The moment-1 map showing the \hon\ \21 velocity field is shown in the right-hand 
panel of Fig.~\ref{fig_mom0overlay}. It shows that the $\hon$ \21 emission exhibits clear signs of rotation. We discuss the implications of the 
$\hon$ \21 observations on the extended $\hon$ disc of the galaxy and the \hon\ \21 absorber in the next section.
%
%
\section{Discussion} 
\label{sec_discussion}
\subsection{Structure and kinematics of the $\hon$ gas associated with \gal}
\label{sec_kinematics}
The double-peaked $\hon$ \21 emission in the Arecibo as well as the GMRT spectra indicate that the $\hon$ disc is rotating at an inclination 
to the line of sight. We find that a simple rotating disc galaxy model following \citet{schulman1994}, with $i$ = 30\textdegree~and maximum 
velocity of 110\,\kms~can reproduce the velocity separation between the two horns of the $\hon$ \21 emission reasonably well (see Fig.~\ref{fig_hiemission}). 
While the galactic disc is inferred to have a zero inclination in the optical bands, the $\hon$ disc inclination from the model is consistent with the 
inclination inferred in the 2MASS infrared K\_s band, which is expected to trace the old stellar population unbiased by dust. By comparing the 
GMRT moment-1 map and the SDSS image of \gal in Fig.~\ref{fig_mom0overlay}, we can see that the $\hon$ \21 emission on the northern (southern) 
part of the galaxy is redshifted (blueshifted) with respect to the systematic velocity. Based on the limited \hon\ \21 data, the kinematical 
major axis of the $\hon$ \21 emission seems to lie along that of the optical and the infrared emission (PA = 160$-$170\textdegree). 
Additionally, we note that there is excess $\hon$ \21 emission in the extreme blue wing of the Arecibo $\hon$ \21 profile ($\sim$ $-$80\,\kms) 
that cannot be explained by a simple rotating disc model. Excess emission in the wings have been explained by high velocity clouds as seen in our 
Galaxy or a warped $\hon$ disc \citep{schulman1994}. Deeper \hon\ \21 emission observations are required to derive the distribution and kinematics 
of \hon\ gas in UGC\,00439, and distinguish between these scenarios.

Further, from Fig.~\ref{fig_hiemission}, the Arecibo $\hon$ \21 emission is asymmetric in amplitude with respect to the systematic velocity of the 
galaxy, with $\sim$20\% more flux in the blue side. This asymmetry is also present to a lesser extent in the GMRT $\hon$ \21 emission. It can be seen 
from Fig.~\ref{fig_mom0overlay}, that the $\hon$ \21 emission on the southern side of the galaxy, which corresponds to the blue peak, seems to be more 
extended than the emission on the northern part, corresponding to the red peak. Asymmetry in the global $\hon$ \21 profile can be traced to asymmetry 
in the $\hon$ gas distribution along with that in the disc kinematics. We find that even in the available UV and infrared images of this galaxy, 
the stellar disc seems to be asymmetric in terms of its spiral structure and light distribution, reflecting the asymmetry evident in the $\hon$ 
gas distribution. \citet{richter1994}, from the frequency of lopsided $\hon$ \21 profiles, derive a lower limit of 50\% for the fraction of galaxies 
with non-circular density distribution. Such lopsidedness may arise out of accretion or merger events in the past.  

\begin{figure*}
\includegraphics[width=0.6\textwidth, angle=270]{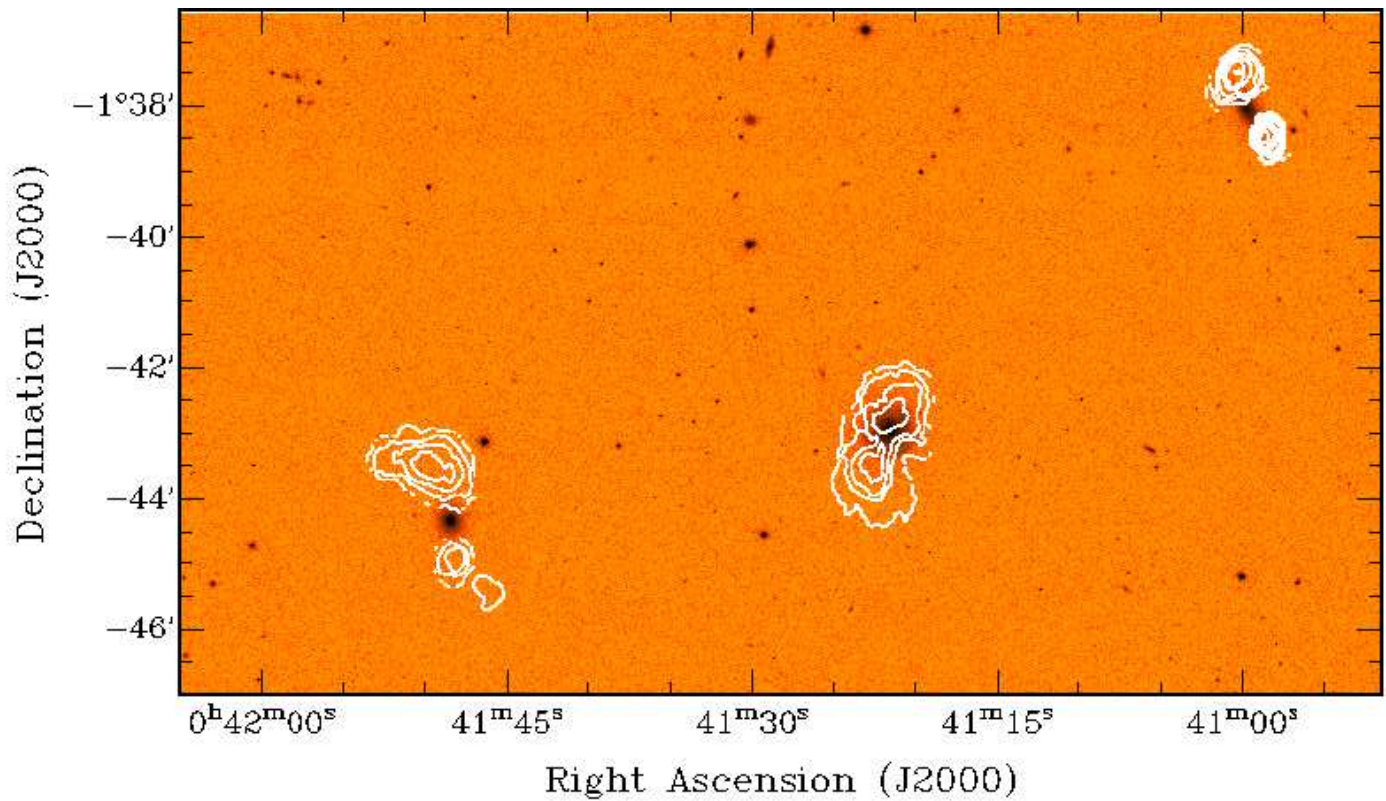}
\caption{SDSS r-band image of the field around \gal overlaid with contours of the GMRT moment-0 map. The contour levels correspond to 
$N$($\hon$) = (0.4, 1.0, 1.6,...) $\times$ 10$^{20}$\,\cms. $\hon$ \21 emission is detected from two other galaxies at 7$'$ ($\sim$150~kpc) 
from \gal in the GMRT field of view: UGC~00435 (J004059.6$-$013802; $z$ = 0.018139; $V_H$ = 5438\,\kms) and CGCG~383$-$072 (J004148.4$-$01441; 
$z$ = 0.018356; $V_H$ = 5503\,\kms).}
\label{fig_groupoverlay}
\end{figure*}

Since UGC\,00439 is part of a galaxy group (see Fig.~\ref{fig_groupoverlay}), this is certainly a possibility. The structure and kinematics of the 
extended $\hon$ disc are known to be highly affected by interaction/merger events, and in few cases the highest density $\hon$ has been observed 
to be beyond the optical discs \citep[e.g.][]{hibbard1996,duc1997,sengupta2013,sengupta2015}. In the GMRT data, we detect $\hon$ \21 emission from two 
other galaxies at $b\sim$150~kpc from \gal in the group (see also Fig.~\ref{fig_groupoverlay}): UGC~00435 (J004059.6$-$013802; $z$ = 0.018139; $V_H$ 
= 5438\,\kms) and CGCG~383$-$072 (J004148.4$-$01441; $z$ = 0.018356; $V_H$ = 5503\,\kms). For the former we measure $\int S dv = 0.52 \pm 0.15$~Jy\,\kms, 
and for the latter $\int S dv = 0.76 \pm 0.20$~Jy\,\kms. UGC~00435 has been studied in $\hon$ \21 emission using Nancay by \citet{theureau2007}, who gives 
$\int S dv = 2.57 \pm 0.47$~Jy\,\kms. The GMRT data recovers $\sim$20\% of the single dish measurement of this galaxy. 

\subsection{Connecting the $\hon$ \21 absorber with the host galaxy}
\label{sec_galaxyabsorber}
The \hon\ \21 absorption occurs at the systematic velocity of the galaxy, as evident from Fig.~\ref{fig_hiemission}, which shows that the peak \hon\ 
\21 absorption towards L2 coincides with the centre of the $\hon$ \21 emission. This indicates that the absorber is located along the kinematic minor 
axis of the $\hon$ disc. Moreover, the absorption is narrow and shows no signatures of outflowing/infalling gas. Hence, the gas detected in \hon\
\21 absorption is most likely to follow the rotation of the extended $\hon$ disc of the galaxy. Recall from Section \ref{sec_21cmextent} that 
the \hon\ \21 absorbing gas shows no systematic velocity gradient over the region ($\sim$2~kpc$^{-2}$) of the southern lobe of J0041$-$0143. But as 
mentioned in Section \ref{sec_metalabsorption} and shown in Fig.~\ref{fig_metalspec}, the metal absorption detected towards the core ($\sim$4~kpc 
north of the southern lobe) is redshifted by $\sim$11\,\kms~with respect to the \hon\ \21 absorption. We estimate the expected velocity shift due to 
rotation with respect to the systematic velocity of the galaxy as $\sim$7\,\kms~at the location of the core (using the maximum rotational 
velocity from the Arecibo $\hon$ \21 emission and the PAs of L2 and C). Hence, both the $\hon$ \21 and the metal absorption could be tracing gas 
that is corotating with the $\hon$ disc. 

In most of the low-$z$ QGPs in which \hon\ \21 absorption has been detected in literature, the radio sightline passes through either the stellar 
disc or the extended $\hon$ disc of the foreground galaxy \citep[$b<$20~kpc,][]{gupta2013,zwaan2015}. On the other hand, impact parameters 
of host galaxy candidates of low-$z$ H$_2$ absorbers have been found to be much further away from the stellar disc \citep[10$\lesssim b$(kpc)$\lesssim$80,][]{muzahid2015}. 
The authors conjecture that the H$_2$ bearing gas may stem from self-shielded, tidally stripped or ejected disc-material in the extended 
galactic halo. In the present case, since the \hon\ \21 absorber is detected away from the optical/UV QSO sightline, we can clearly see that 
there is no indication of star formation (i.e. no stellar light) at the location of the \hon\ \21 absorber in the available optical and UV 
images (see e.g. the SDSS r-band image in Fig.~\ref{fig_mom0overlay}). We estimate m$_r$ = 21.9 and m$_{NUV}$ = 22.5 in a circular 
aperture of 5$''$ placed over the southern lobe in the SDSS r-band and GALEX near-UV images respectively. This suggests that the absorption 
is unlikely to arise from the ISM of a low-luminosity dwarf galaxy at that location. Hence, this supports our above conjecture that we 
are most likely tracing clumps of corotating cold gas in the extended $\hon$ disc of the galaxy.

\begin{figure}
\includegraphics[width=0.38\textwidth, angle=90]{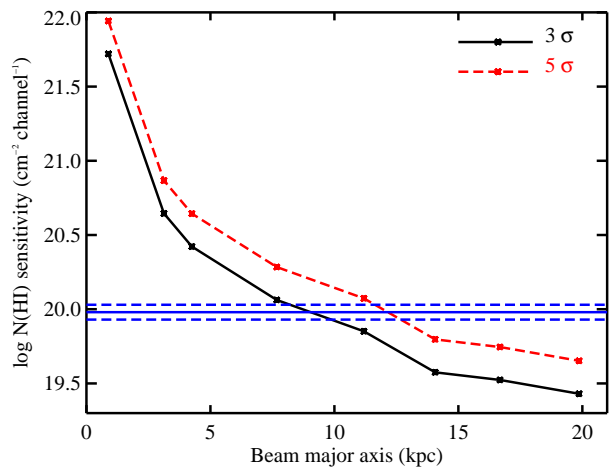}
\caption{The 3$\sigma$ (5$\sigma$) $N$($\hon$) sensitivity per channel (7.1\,\kms) at different spatial resolutions, in the GMRT 
$\hon$ \21 emission data set, is shown as a solid (dashed) line. The horizontal solid line marks the log~$N$($\hon$) corresponding   
to the optical depth detected towards L2 for a 100~K gas with $f_c$ = 1 (see Table \ref{tab:radioparameters}), and the dashed horizontal 
lines show the 1$\sigma$ errors associated with it.}
\label{fig_sensitivity}
\end{figure}

We note that better sensitivity $\hon$ \21 emission maps of the extended $\hon$ gas distribution around the galaxy are required to quantify 
the physical conditions in the absorbing gas. However, we can draw some simple estimates of the physical conditions in the absorbing 
gas using the $N$($\hon$) sensitivity of the GMRT data across different spatial scales. As described in Section \ref{sec_21cmemission}, 
we constructed image cubes of different spatial resolutions in order to detect $\hon$ \21 emission. In Fig.~\ref{fig_sensitivity}, we show 
the $N$($\hon$) sensitivity (at 3 and 5 $\sigma$ significance) per velocity channel (7.1\,\kms) for different spatial resolutions, estimated 
using the rms in line-free channels in these image cubes. Note that the \hon\ \21 absorption is spread over only one channel at the velocity 
resolution in these cubes. The horizontal line marks the log~$N$($\hon$) we expect to be associated with the absorbing gas towards L2 if 
it has $T_s$ = 100~K and $f_c$ = 1 (see Table \ref{tab:radioparameters}). $\hon$ \21 emission is not detected at the location of the \hon\ \21 
absorption in the image cubes with a resolution of $\ge$10~kpc, where we have sufficient sensitivity to detect up to typical $N$($\hon$) seen 
in sub-damped Lyman-$\alpha$ systems \citep[$N$($\hon$) $\ge$ 10$^{19}$\,\cms; see][]{wolfe1982, wolfe2005, peroux2005}.
This implies that either the absorbing gas, spread over $\ge$10~kpc, has temperature less than 100~K, or that the cold $\sim$100~K gas is more 
compact than $\sim$10~kpc. In Section \ref{sec_21cmextent}, we saw that the absorbing gas extends up to at least 1~kpc and may extend up to $\sim$4~kpc 
with optical depth varying by a factor of $\gtrsim$2. The present data does not have sufficient sensitivity to detect cold 100~K gas of size $\sim$4~kpc. 
However, from the non-detection of $\hon$ \21 emission over this scale we get a limit on the temperature of the absorbing gas as $\lesssim$300~K. 

The fact that the \hon\ \21 optical depth falls by a factor of $\gtrsim$7 over $\sim$7~kpc at similar impact parameters from the galaxy 
could be a manifestation of patchy distribution of cold gas in the extended \hon\ disc, with a varying covering factor at different spatial locations.
This variation in optical depth at similar radius from the galaxy should be taken into account when interpreting the covering factor of \hon\ 
\21 absorbers around low-$z$ galaxies.
%
%
\section{Summary}  
\label{sec_summary} 
This paper presents GMRT $\hon$ \21 absorption and emission observations of the QGP, J0041$-$0143$/$UGC~00439. The background radio 
source consists of a core component and two lobes, all at a projected separation of $\sim$25~kpc from the $z$ = 0.01769 galaxy, UGC~00439. 
We detect \hon\ \21 absorption towards the southern lobe but not towards the core ($\sim$4~kpc north of southern lobe) or the northern 
lobe ($\sim$7~kpc north of southern lobe). The \hon\ \21 absorption towards the continuum peak of the southern lobe corresponds to an 
integrated optical depth of 0.52 $\pm$ 0.07\,\kms~and a FWHM of 8 $\pm$ 1\,\kms. From channel maps of the absorption and spatially
resolved extraction of \hon\ \21 spectra over the southern lobe, we surmise that the absorbing gas is likely to have sub-kpc structures
and turbulent motion of the order of few\,\kms, extending over a $\sim$2~kpc$^2$ region. The absorbing gas may extend up to the core, 
which is cospatial with the optical QSO and towards which we detect $\cat$ and $\nao$ absorption in the Keck-HIRES spectrum. 

The GMRT $\hon$ \21 emission associated with UGC~00439 recovers $\sim$20\% of the total flux density measured by Arecibo. The Arecibo and 
GMRT $\hon$ \21 observations of this galaxy suggest that its $\hon$ disc is lopsided, which may arise out of interaction/merger events. 
Detection of $\hon$ \21 emission from two other neighbouring galaxies, $\sim$150~kpc away from UGC~00439, in the GMRT data, gives 
credence to the idea that some of the $\hon$ gas in \gal may have been stripped out of its regular rotating disc due to tidal interactions with 
neighbouring galaxies. From the $N$($\hon$) sensitivity of the GMRT observations at different spatial scales, we estimate that if the 
absorber had a spatial extent of $\sim$4~kpc, its temperature would be $\lesssim$300~K. Further, the $\hon$ \21 absorption arises at the 
systematic redshift of the galaxy, with no stellar light detected at its location, implying that it is most likely tracing clumpy gas 
corotating with the $\hon$ disc of the galaxy. 

The GMRT observations reveal kpc to sub-kpc structures in the $\hon$ gas around UGC~00439. However, it still leaves some questions 
regarding the nature of the $\hon$ gas unanswered. A higher sensitivity $\hon$ \21 emission map is required to study the $\hon$ gas 
distribution at the location of the background radio source. Further, subarcsecond-scale radio observations of \sys will help in 
quantifying the covering factor and parsec-scale structures in the absorbing gas. In addition, two-dimensional optical spectra of 
\gal along with deeper $\hon$ \21 observations would help to better compare the distribution and kinematics of the stellar population 
and the $\hon$ gas. Lastly, higher sensitivity $\hon$ \21 emission observations would help to map any large-scale diffuse $\hon$ gas 
that may exist as a result of tidal interactions/mergers between the nearby galaxies.

Hence, while the results presented in this paper help in characterizing the $\hon$ gas around UGC~00439, they also motivate future 
observations of this system. In addition, the results presented in this paper along with our ongoing study of \hon\ \21 absorption in 
the discs/haloes of low-$z$ galaxies \citep[][Dutta et al. in preparation]{gupta2010,gupta2013,srianand2013,srianand2015}, are 
expected to establish the elusive link between \hon\ \21 absorbers and host galaxy properties, and play a vital role in understanding 
the nature of high-$z$ \hon\ \21 absorbers \citep[][]{briggs1983,kanekar2003,Curran05,gupta2009,gupta2012,srianand2012,kanekar2014}, 
as well as the large number of \hon\ \21 absorbers expected to be detected from upcoming blind surveys (e.g. MeerKAT Absorption Line Survey) 
using Square Kilometre Array pathfinders and precursors. \\
%
%

\noindent \textbf{ACKNOWLEDGEMENTS} \newline

\noindent 
We thank the anonymous referee for his/her useful comments.
We thank the GMRT staff for their help during the observations. GMRT is run by the National Centre for Radio Astrophysics of the
Tata Institute of Fundamental Research. Some of the data presented in this work were obtained from the Keck Observatory Database 
of Ionized Absorbers toward Quasars (KODIAQ), which was funded through NASA ADAP grant NNX10AE84G.
%
%
\def\aj{AJ}%
\def\actaa{Acta Astron.}%
\def\araa{ARA\&A}%
\def\apj{ApJ}%
\def\apjl{ApJ}%
\def\apjs{ApJS}%
\def\ao{Appl.~Opt.}%
\def\apss{Ap\&SS}%
\def\aap{A\&A}%
\def\aapr{A\&A~Rev.}%
\def\aaps{A\&AS}%
\def\azh{A$Z$h}%
\def\baas{BAAS}%
\def\bac{Bull. astr. Inst. Czechosl.}%
\def\caa{Chinese Astron. Astrophys.}%
\def\cjaa{Chinese J. Astron. Astrophys.}%
\def\icarus{Icarus}%
\def\jcap{J. Cosmology Astropart. Phys.}%
\def\jrasc{JRASC}%
\def\mnras{MNRAS}%
\def\memras{MmRAS}%
\def\na{New A}%
\def\nar{New A Rev.}%
\def\pasa{PASA}%
\def\pra{Phys.~Rev.~A}%
\def\prb{Phys.~Rev.~B}%
\def\prc{Phys.~Rev.~C}%
\def\prd{Phys.~Rev.~D}%
\def\pre{Phys.~Rev.~E}%
\def\prl{Phys.~Rev.~Lett.}%
\def\pasp{PASP}%
\def\pasj{PASJ}%
\def\qjras{QJRAS}%
\def\rmxaa{Rev. Mexicana Astron. Astrofis.}%
\def\skytel{S\&T}%
\def\solphys{Sol.~Phys.}%
\def\sovast{Soviet~Ast.}%
\def\ssr{Space~Sci.~Rev.}%
\def\zap{$Z$Ap}%
\def\nat{Nature}%
\def\iaucirc{IAU~Circ.}%
\def\aplett{Astrophys.~Lett.}%
\def\apspr{Astrophys.~Space~Phys.~Res.}%
\def\bain{Bull.~Astron.~Inst.~Netherlands}%
\def\fcp{Fund.~Cosmic~Phys.}%
\def\gca{Geochim.~Cosmochim.~Acta}%
\def\grl{Geophys.~Res.~Lett.}%
\def\jcp{J.~Chem.~Phys.}%
\def\jgr{J.~Geophys.~Res.}%
\def\jqsrt{J.~Quant.~Spec.~Radiat.~Transf.}%
\def\memsai{Mem.~Soc.~Astron.~Italiana}%
\def\nphysa{Nucl.~Phys.~A}%
\def\physrep{Phys.~Rep.}%
\def\physscr{Phys.~Scr}%
\def\planss{Planet.~Space~Sci.}%
\def\procspie{Proc.~SPIE}%
\let\astap=\aap
\let\apjlett=\apjl
\let\apjsupp=\apjs
\let\applopt=\ao
\bibliographystyle{mn}
\bibliography{mybib}
\end{document}